\documentclass[12pt]{article}
\usepackage{graphicx,rotating}
\newcommand{\HRule}{\rule{0.4\linewidth}{0.3mm}}
\newcommand{\mlab}[1]%
    {\mbox{}\marginpar{\raggedright\hspace{0pt}\footnotesize #1}}

\newcommand{\jpsi}{J/$\psi$}
\newcommand{\psip}{$\psi^\prime$}

\newcommand{\psidy}{(\jpsi)\,/\,Drell-Yan}
\newcommand{\et}{$E_{\rm T}$}
\newcommand{\pt}{$p_{\rm T}$}
\newcommand{\ezdc}{$E_{\rm ZDC}$}
\newcommand{\npart}{$N_{\rm part}$}
\newcommand{\dnde}{$N_{\rm ch}$}
\newcommand{\dd}{{\rm d}}

\begin{document}

\thispagestyle{empty}
\baselineskip=14pt
\parskip 0pt plus 5pt

\begin{center}
{\large EUROPEAN LABORATORY FOR PARTICLE PHYSICS}
\end{center}

\bigskip
\begin{flushright}
CERN--PH--EP\,/\,2004--052\\
September 15, 2004
\end{flushright}

\bigskip\bigskip
\begin{center}
\Large\bf{A new measurement of \jpsi\ suppression\\ 
in Pb-Pb collisions at 158~GeV per nucleon} 
\end{center}

\bigskip\bigskip
\begin{center}
\emph{\large NA50 collaboration}
\end{center}
\bigskip\bigskip

\begin{center}
\textbf{Abstract}
\end{center}

\bigskip
\begingroup
\leftskip=0.4cm
\rightskip=0.4cm
\parindent=0.pt
\small

We present a new measurement of \jpsi\ production in
\mbox{Pb-Pb} collisions at 158~GeV/nucleon, from the data sample
collected in year 2000 by the NA50 Collaboration, under improved
experimental conditions with respect to previous years. With the target system placed in vacuum, 
the setup was better adapted to study, in particular, the most 
peripheral nuclear collisions with unprecedented accuracy. 
The analysis of this data sample shows that
the \psidy\ cross-sections ratio measured in the most peripheral
\mbox{Pb-Pb} interactions is in good agreement with the nuclear
absorption pattern extrapolated from the studies of 
proton-nucleus collisions. 
Furthermore, this new measurement confirms our previous observation 
that the \psidy\ cross-sections ratio departs from the normal
nuclear absorption pattern for semi-central \mbox{Pb-Pb} collisions
and that this ratio persistently decreases up to the most central collisions.

\bigskip\bigskip

\begin{center}
\emph{(submitted to Eur. Phys. J. C - REVISED VERSION)}
\end{center}

\newpage

\begin{center}
B.~Alessandro$^{10}$,
C.~Alexa$^{3}$,
R.~Arnaldi$^{10}$,
M.~Atayan$^{12}$,
S.~Beol\`e$^{10}$,
V.~Boldea$^{3}$,
P.~Bordalo$^{6,a}$,
G.~Borges$^{6}$,
C.~Castanier$^{2}$,
J.~Castor$^{2}$,
B.~Chaurand$^{9}$,
B.~Cheynis$^{11}$,
E.~Chiavassa$^{10}$,
C.~Cical\`o$^{4}$,
M.P.~Comets$^{8}$,
S.~Constantinescu$^{3}$,
P.~Cortese$^{1}$,
A.~De~Falco$^{4}$,
N.~De~Marco$^{10}$,
G.~Dellacasa$^{1}$,
A.~Devaux$^{2}$,
S.~Dita$^{3}$,
O.~Drapier$^{9}$,
J.~Fargeix$^{2}$,
P.~Force$^{2}$,
M.~Gallio$^{10}$,
C.~Gerschel$^{8}$,
P.~Giubellino$^{10}$,
M.B.~Golubeva$^{7}$,
M.~Gonin$^{9}$,
A.~Grigoryan$^{12}$,
S.~Grigoryan$^{12}$,
F.F.~Guber$^{7}$,
A.~Guichard$^{11}$,
H.~Gulkanyan$^{12}$,
M.~Idzik$^{10,b}$,
D.~Jouan$^{8}$,
T.L.~Karavicheva$^{7}$,
L.~Kluberg$^{9}$,
A.B.~Kurepin$^{7}$,
Y.~Le~Bornec$^{8}$,
C.~Louren\c co$^{5}$,
M.~Mac~Cormick$^{8}$,
P.~Macciotta$^{4}$,
A.~Marzari-Chiesa$^{10}$,
M.~Masera$^{10}$,
A.~Masoni$^{4}$,
M.~Monteno$^{10}$,
A.~Musso$^{10}$,
P.~Petiau$^{9}$,
A.~Piccotti$^{10}$,
J.R.~Pizzi$^{11}$,
F.~Prino$^{10}$,
G.~Puddu$^{4}$,
C.~Quintans$^{6}$,
L.~Ramello$^{1}$,
S.~Ramos$^{6,a}$,
L.~Riccati$^{10}$,
A.~Romana$^{9}$,
H.~Santos$^{6}$,
P.~Saturnini$^{2}$,
E.~Scomparin$^{10}$,
S.~Serci$^{4}$,
R.~Shahoyan$^{6,c}$,
F.~Sigaudo$^{10}$,
M.~Sitta$^{1}$,
P.~Sonderegger$^{5,a}$,
X.~Tarrago$^{8}$,
N.S.~Topilskaya$^{7}$,
G.L.~Usai$^{4}$,
E.~Vercellin$^{10}$,
L.~Villatte$^{8}$,
N.~Willis$^{8}$,
T.~Wu$^{8}$.
\end{center}

\setlength{\parindent}{0mm}
\setlength{\parskip}{0mm}
\small
\HRule\\
\begin{flushleft}
$^{~1}$ Universit\`a del Piemonte Orientale, Alessandria and INFN-Torino, Italy  \\
$^{~2}$ LPC, Univ. Blaise Pascal and CNRS-IN2P3, Aubi\`ere, France\\
$^{~3}$ IFA, Bucharest, Romania\\
$^{~4}$ Universit\`a di Cagliari/INFN, Cagliari, Italy\\
$^{~5}$ CERN, Geneva, Switzerland\\
$^{~6}$ LIP, Lisbon, Portugal\\
$^{~7}$ INR, Moscow, Russia\\
$^{~8}$ IPN, Univ. de Paris-Sud and CNRS-IN2P3, Orsay, France\\
$^{~9}$ LLR, Ecole Polytechnique and CNRS-IN2P3, Palaiseau, France\\
$^{10}$ Universit\`a di Torino/INFN, Torino, Italy\\
$^{11}$ IPN, Univ. Claude Bernard Lyon-I and CNRS-IN2P3, Villeurbanne, France\\
$^{12}$ YerPhI, Yerevan, Armenia\\

a) also at IST, Universidade T\'ecnica de Lisboa, Lisbon, Portugal\\
b) also at Faculty of Physics and Nuclear Techniques, AGH University of
Science and Technology, Cracow, Poland\\
c) now at CFTP, IST, Lisbon, Portugal\\

\end{flushleft}
\endgroup

\newpage
\pagenumbering{arabic}
\setcounter{page}{1}

 
\section{Introduction}

The suppression of the \jpsi\ yield in heavy ion collisions, predicted
by Matsui and Satz~\cite{Mat86}, is commonly considered as one of the most interesting signals
of the formation of a deconfined state of quarks and gluons
in high-energy heavy-ion collisions.
The detection of \jpsi\ and \psip\ mesons through their leptonic decay to a pair of muons
is particularly interesting since muons are not affected by the strong
interactions at play in the later stages of the collision evolution,
when the light hadrons are formed.

The NA50 experiment is a high luminosity fixed target experiment at CERN, essentially 
dedicated to the study of dimuon production in Pb-Pb collisions at
158~GeV per nucleon.
The analysis of the Pb-Pb data collected by NA50 in year 1995
showed~\cite{Abr97a,Abr97b} that the \jpsi\ production
yield, with respect to the production of Drell-Yan dimuons, is
(``anomalously'') suppressed with respect to
the ``normal nuclear absorption'' pattern derived from
measurements done in interactions induced by
protons or light nuclei~\cite{Abr98,Abr99a,Abr99b}.
This ``integrated'' result was complemented by detailed
studies of the \jpsi\ suppression pattern as a function of the
centrality of the collision~\cite{Abr99,Abr00,Abr01}, with data collected
in the years 1996 and 1998, which indicated that the extra suppression
sets in for semi-central collisions and suggested that the departure from the
normal absorption curve was setting in over a narrow range of centrality
values.

A more detailed analysis
of the Pb-Pb data revealed that peripheral
interactions could be contaminated by Pb-air interactions, especially
in the multi-target configuration used in the data taking
period of 1996.
We devoted the year 2000 data taking period to investigate further 
whether Pb-Pb peripheral collisions 
were really compatible with the results obtained from lighter collision systems, and to collect 
more statistics on several \mbox{p-A} systems, in order to establish a more precise normal absorption 
curve.
It was also realized that,
although the pattern describing their dependence as a function of the 
centrality of the collision remains essentially unaffected,
the numerical
values of the ratio of cross-sections $B_{\mu\mu}\sigma({\rm J}/\psi)/\sigma(DY)$
are somewhat sensitive to the specific parametrization 
used for the Parton Distribution Functions (PDFs).
This happens because different PDFs give slightly different shapes
for the Drell-Yan dimuon mass distribution
and, therefore, as a result of the constraint imposed by the
data in the high mass region, lead to a different Drell-Yan yield 
in the region $ 2.9 < M < 4.5 \ {\rm GeV}/c^2$. 
Since the GRV~92~LO~\cite{GRVLO} PDFs were used for the analyses of the S-U data
and of the 1995 Pb-Pb data, while
the MRS~A (Low Q$^2$)~\cite{MRS43} PDFs (which take into account the isospin
asymmetry in the quark sea, $\overline{u}\neq\overline{d}$, as
determined by the NA51 experiment~\cite{NA51}) were taken for later data
analyses, there is a systematic discrepancy between the corresponding
results.
To overcome this small inconsistency, 
all the data referred to in this paper, both the Pb-Pb event samples and
the lighter interacting systems used as reference, have been analyzed
with a single set of PDFs.  We have chosen 
the GRV~94~LO~\cite{GRV94}
PDFs, to take into account the quark sea asymmetry and
to be consistent with the fact that 
the Monte Carlo event generators that we use are based on leading order calculations.
Finally, we also report results normalized to the Drell-Yan yield 
in the muon pair mass range \mbox{$ 4.2 < M < 7.0 \ {\rm GeV}/c^2$}, which have the
advantage of being essentially insensitive to the PDFs chosen for the analysis.  

In summary, this paper presents the measurement of the \psidy\ cross-sections ratio, as a function of 
collision centrality, from data collected in year 2000, using our most recent data selection 
and analysis procedures.
Preliminary results of this analysis have been previously presented in Ref.~\cite{Ram03}.


\section{Experimental setup, event reconstruction and data selection}

\begin{figure}[htb]
\centering
\begin{turn}{-90}
\resizebox{0.7\textwidth}{!}{%
\includegraphics*[bb=242 0 595 486]{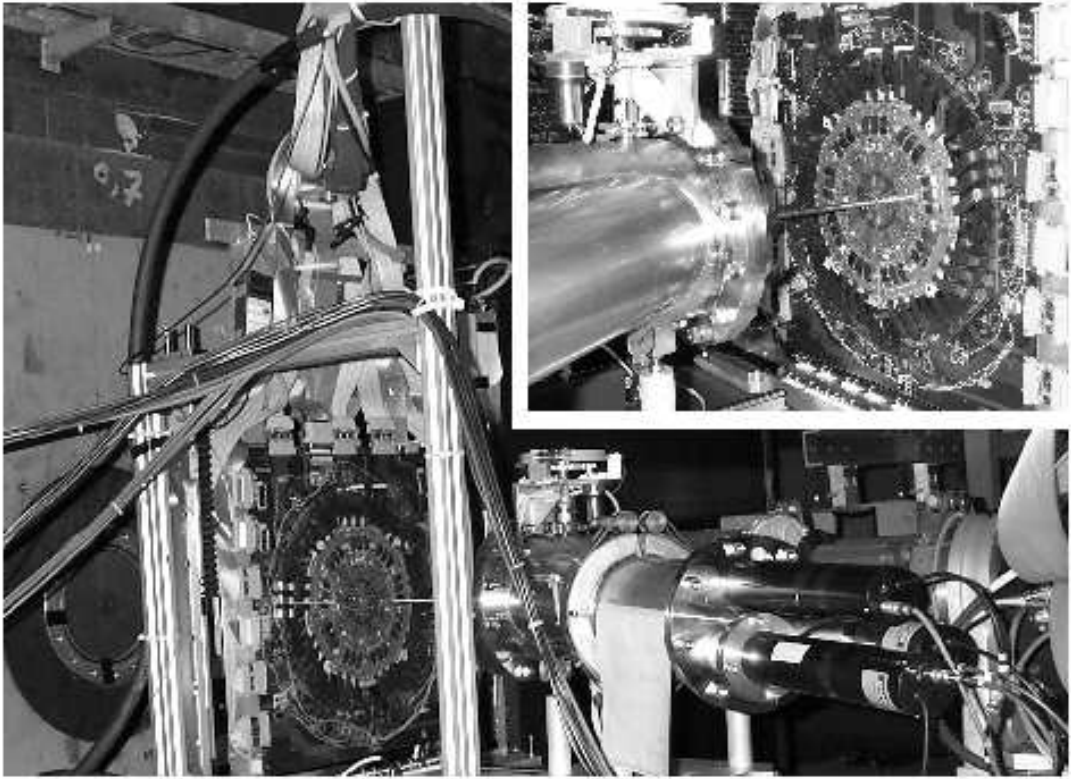}}
\end{turn}
\caption{The NA50 target region (beam entering from bottom right), including the target 
vacuum tank, the Multiplicity Detector and the upstream face of the carbon absorber;
the Electromagnetic Calorimeter (not shown) with the BeO preabsorber in its
center is inserted between the MD and the carbon absorber during data taking. 
The inset shows part of the target vacuum tank and the upstream face of the Multiplicity 
Detector's first plane (beam entering from left). During data taking conditions
the target system is moved down-stream, almost touching the MD.}
\label{fig:target}
\end{figure}

The experimental setup used by NA50 in the year 2000 included
important improvements with respect to the one used in the previous
Pb-Pb data taking periods of 1995, 1996 and 1998.
The main change was the introduction of a new target system under
vacuum~\cite{Cas03} (see Fig.~\ref{fig:target}), which allowed a better rejection 
of out-of-target interactions and, in particular, of \mbox{Pb-air} interactions. 
For data selection, a new target
identification algorithm was used, based on the silicon Multiplicity
Detector~\cite{Ale02} instead of the previous one, based on a system
of quartz blades~\cite{Bel97}. Furthermore, in order to identify
almost simultaneous multiple interactions, a new method was developed 
based on the shape analysis of the signals of the
Electromagnetic Calorimeter.  
Except for the new target system, the setup is described in detail in 
Refs.~\cite{Abr97a,Arn98}.  Its main features are recalled hereafter.

The NA50 apparatus consists essentially of a set of beam and anti-halo
counters, three centrality detectors, and a muon spectrometer.  The
intensity of the incoming ion beam is counted by a Beam Hodoscope (BH)
made of 16 quartz slabs. Interactions occurring upstream of the target
are vetoed by dedicated scintillator counters.  Interactions in the
target can be tagged by two quartz counters or by the two planes of silicon
micro-strips of the Multiplicity Detector (MD), depicted in
Fig.~\ref{fig:target}, which also count the charged particles
produced in the angular range $1.9<\eta<4.2$.  The other two
centrality detectors are the Electromagnetic Calorimeter (EMC), which
integrates the flux of neutral transverse energy \et\ in the angular domain
$1.1<\eta<2.3$, and the Zero-Degree Calorimeter (ZDC), which measures \ezdc,
the forward energy carried by the beam spectator nucleons ($\eta > 6.3$).  The muon
spectrometer starts with a hadron absorber, made of beryllium oxide, carbon and
iron, followed by two sets of multi-wire proportional chambers, 
located upstream and downstream of an air-core toroidal magnet.  
The dimuon trigger is provided
by 4 scintillator hodoscopes. 
A minimum bias (MB) 
trigger is defined by a minimal energy deposited in the ZDC
and a ``beam'' trigger is provided by the BH, leading to a sample of
events only constrained by the detection of an identified incident 
Pb ion.  
Data in year 2000
were taken with the 158~GeV/nucleon beam at 1--1.4$\cdot 10^7$~ions/s,
over 5~s bursts, and with a 4.0~mm thick Pb target, corresponding to 10\,\% of
an interaction length (interaction probability: 9.6\,\%).
In 35 days, we collected 135 million dimuon triggers on tape (plus
additional triggers without target or at lower beam intensities) which
led to 64 million reconstructed dimuons, of which 720\,000 opposite-sign pairs 
in the \jpsi\ mass region (2.9--3.3~GeV/$c^2$).  After data
selection and background subtraction (see below) the number of useful
dimuons in the \jpsi\ mass region was between 100\,000 and 130\,000
(depending on the particular definition of selection cuts).

\begin{figure}[htb]
\centering
\resizebox{0.65\textwidth}{!}{\includegraphics*{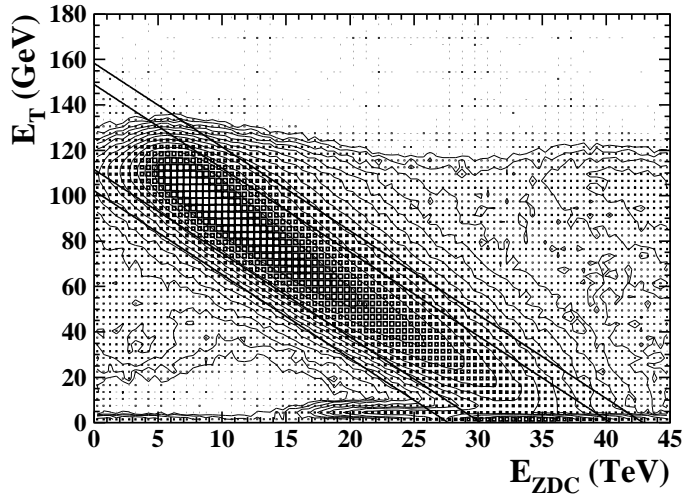}}
\caption{Correlation between transverse and forward energy, showing $\pm 2 \sigma$ and $\pm 3 \sigma$ 
diagonal bands. Contours connect pixels of equal event density, ranging from 20 
(external one) to 5024 (closest to the highest density), scaled with a constant factor.}
\label{fig:diag}
\end{figure}

The events were processed offline through a technically improved version of the software,
featuring a higher track reconstruction efficiency~\cite{Ale04,Sha01}
with respect to the version used for data taking periods before year
2000.  The data selection proceeded as follows.  First of all,
parasitic interactions of the incident Pb ion in the BH were rejected
using auxiliary scintillator counters. Then, multiple interactions were
rejected by a temporal analysis of the signal in the EMC, allowing us
to retain events where either one or two incident ions were detected in the
BH, within a given time window, 
but only one ion interacted in the target.  
Some residual interaction pile-up events were further 
rejected by a diagonal band cut on the \et--\ezdc\
correlation, using the method described in Ref.~\cite{Qui02} (Fig.~\ref{fig:diag}).

The location of the primary interaction is determined by requiring 
the appropriate correlation between hits on the first and second planes of the MD.
This method works for
$E_{\rm T}>3$~GeV, and reaches full efficiency at $E_{\rm T}\simeq
15$~GeV (see also Ref.~\cite{Ale02}).
It retains for further analysis significantly more peripheral events
than the traditional target identification algorithm, based on quartz Cherenkov
counters surrounding the target.

Muons originating downstream from the target (e.g.\ due to collisions in the BeO preabsorber or in
the ZDC) are rejected by a cut on $p \cdot D_{\rm Targ}$, where $p$ is the momentum
and $D_{\rm Targ}$ is the transverse distance between
the extrapolated muon track at the target position and the beam axis.
For the analysis of this paper, which is focused on high mass dimuons, 
the level of this ``muon target cut'' was set
at 2\,\% of $\chi^2$ probability for each muon (see Ref.~\cite{San04} for details). 
We have verified 
that our results are stable when we increase the level of the cut up to 10\,\%, while 
of course the number of accepted dimuons is consequently reduced.
The rather low level of the cut with respect to previous data taking periods is allowed by the
improved experimental conditions in the year 2000 run, in particular the
vacuum pipe extending inside the BeO preabsorber.
We reach a quite effective rejection of the
out-of-target events, which are especially numerous in the most peripheral
centrality bin at masses 
around 2.5~GeV/$c^2$ (see also Ref.~\cite{Ram03}).

\begin{table}[ht]
\centering
\begin{tabular}{|c|cc|cc|cc|cc|}
\hline
Class 
& \multicolumn{2}{|c|}{$E_{\rm T}$ (GeV)} 
& \multicolumn{2}{|c|}{$N_{\rm part}$}
& \multicolumn{2}{|c|}{$b$ (fm)}
& \multicolumn{2}{|c|}{$L$ (fm)} \\
& range & average 
& average & rms 
& average & rms 
& average & rms \\
\hline
 $1_T$ &   3--15  &  10.6 &  34 & 13 & 11.8 & 0.7 & 4.44 & 0.72 \\ %
 $2_T$ &  15--25  &  20.4 &  70 & 13 & 10.3 & 0.5 & 5.94 & 0.42 \\ %
 $3_T$ &  25--35  &  30.3 & 104 & 14 &  9.2 & 0.4 & 6.84 & 0.33 \\ %
 $4_T$ &  35--45  &  40.2 & 138 & 16 &  8.2 & 0.4 & 7.51 & 0.27 \\ %
 $5_T$ &  45--55  &  50.2 & 172 & 17 &  7.3 & 0.4 & 8.02 & 0.23 \\ %
 $6_T$ &  55--65  &  60.1 & 206 & 18 &  6.5 & 0.4 & 8.43 & 0.20 \\ %
 $7_T$ &  65--75  &  70.1 & 240 & 19 &  5.6 & 0.5 & 8.76 & 0.17 \\ %
 $8_T$ &  75--85  &  80.1 & 274 & 20 &  4.8 & 0.5 & 9.02 & 0.15 \\ %
 $9_T$ &  85--95  &  90.1 & 308 & 21 &  3.9 & 0.6 & 9.22 & 0.12 \\ %
$10_T$ &  95--105 & 100.0 & 341 & 20 &  2.8 & 0.7 & 9.38 & 0.11 \\ %
$11_T$ & 105--150 & 111.5 & 367 & 15 &  1.7 & 0.7 & 9.48 & 0.06 \\ %
\hline
\end{tabular}
\caption{Centrality classes based on the transverse energy
  measurement. For each class we list the $E_{\rm T}$ range and
  average, together with the average and rms values of $N_{\rm part}$,
  $b$ and $L$.}
\label{tab:centralet}
\end{table}

%
%
%
\begin{table}[ht]
\centering
\begin{tabular}{|c|cc|cc|cc|cc|}
\hline
Class 
& \multicolumn{2}{|c|}{$E_{\rm ZDC}$ (TeV)} 
& \multicolumn{2}{|c|}{$N_{\rm part}$}
& \multicolumn{2}{|c|}{$b$ (fm)}
& \multicolumn{2}{|c|}{$L$ (fm)} \\
& range & average 
& average & rms 
& average & rms 
& average & rms \\
\hline                                                            
$1_F$ & 31--36 & 33.0 &  44 & 27 & 11.6  & 1.4  &  4.70 & 1.27 \\ 
$2_F$ & 27--31 & 28.9 &  76 & 35 & 10.3  & 1.3  &  5.89 & 1.18 \\ 
$3_F$ & 23--27 & 25.0 & 119 & 39 &  8.8  & 1.2  &  7.04 & 0.88 \\ 
$4_F$ & 19--23 & 21.0 & 170 & 39 &  7.4  & 1.0  &  7.92 & 0.59 \\ 
$5_F$ & 15--19 & 17.0 & 223 & 38 &  6.1  & 0.9  &  8.56 & 0.39 \\ 
$6_F$ & 11--15 & 13.0 & 276 & 35 &  4.7  & 0.9  &  9.01 & 0.26 \\ 
$7_F$ &  7--11 &  9.1 & 330 & 31 &  3.2  & 1.0  &  9.32 & 0.16 \\ 
$8_F$ &   0--7 &  5.7 & 365 & 19 &  1.8  & 0.9  &  9.47 & 0.09 \\ 
\hline
\end{tabular}
\caption{Centrality classes based on the forward energy measurement.}
\label{tab:centralezdc}
\end{table}

%
%
\begin{table}[ht]
\centering
\begin{tabular}{|c|cc|cc|cc|cc|}
\hline
Class 
& \multicolumn{2}{|c|}{$\dd N_{ch}/\dd\eta$} 
& \multicolumn{2}{|c|}{$N_{\rm part}$}
& \multicolumn{2}{|c|}{$b$ (fm)}
& \multicolumn{2}{|c|}{$L$ (fm)} \\
& range & average 
& average & rms 
& average & rms 
& average & rms \\
\hline
$1_N$ & 1--41   &  22.4  &  39    & 21    & 11.7  & 1.0  &  4.58 & 0.97 \\
$2_N$ & 41--81  &  62.0  &  70    & 25    & 10.4  & 0.9  &  5.83 & 0.78 \\
$3_N$ & 81--120 &  101.2  & 101   & 29    &  9.3  & 0.9  &  6.70 & 0.66 \\
$4_N$ & 120--173 & 147.2  & 138   & 34    &  8.3  & 0.9  &  7.44 & 0.57 \\
$5_N$ & 173--226 & 200.0  & 179   & 38    &  7.2  & 1.0  &  8.06 & 0.48 \\
$6_N$ & 226--279 & 253.0  & 220   & 41    &  6.2  & 1.0  &  8.53 & 0.40 \\
$7_N$ & 279--332 & 305.7  & 260   & 43    &  5.1  & 1.1  &  8.88 & 0.32 \\
$8_N$ & 332--385 & 358.1  & 297   & 41    &  4.1  & 1.2  &  9.13 & 0.25 \\
$9_N$ & 385--438 & 410.1  & 327   & 36    &  3.2  & 1.2  &  9.30 & 0.18 \\
$10_N$& 438--835 & 487.9  & 352   & 27    &  2.3  & 1.1  &  9.42 & 0.13 \\
\hline
\end{tabular}
\caption{Centrality classes based on the measurement of the charged
  particle multiplicity.}
\label{tab:centraldnde}
\end{table}

Once the event selection is done, we can proceed with the definition of centrality classes.
For this purpose we have used our three centrality detectors,
with the corresponding variables: neutral transverse energy \et, forward energy \ezdc\ 
and charged particle multiplicity per unit of pseudo-rapidity at
mid-rapidity, $({\rm d}N_{\rm ch}/{\rm d}\eta)|_{\rm max}$, noted as \dnde\ in the
following for simplicity of notation~\cite{DNDETA1}.
While \et\ and \dnde\ are more directly
correlated with the energy density of the collision, \ezdc\ is a good
estimator of the 
number of participant nucleons~\cite{Abr01}.
The centrality classes used in this paper are listed in Tables~\ref{tab:centralet},
 \ref{tab:centralezdc} and \ref{tab:centraldnde}.
The bins are equidistant in the measured centrality estimator being used, except for 
the first (or first three in the case of \dnde) and last bin.
These tables include
the corresponding average and rms values of the number of participant
nucleons, $N_{\rm part}$, of the 
impact parameter, $b$, and of the average length of nuclear matter
traversed by the (pre-resonant) charmonium state, $L$,
all of them evaluated through a detailed Glauber calculation (see Ref.~\cite{Kha97} for the Glauber formalism 
and Ref.~\cite{DNDETA2} for an example of such an evaluation). 
For the Pb nuclear density we have used a 2-parameter Fermi distribution
with a half-density radius of 6.624~fm and a diffuseness parameter of
0.549~fm (see Ref.~\cite{deJ74} and reference Ja73 therein).
The parameters quoted above only describe the proton distribution
inside the Pb nucleus, and it is well known that neutrons are
distributed differently. Since our reference, the
Drell-Yan process, is isospin-dependent, we have used a recent experimental measurement~\cite{Trz01} 
of the ``neutron halo'' in heavy nuclei to define a 2-parameter Fermi distribution for neutrons, with 
the same half-density radius as for protons and a diffuseness parameter of 0.667~fm (for more 
details see Ref.~\cite{NA50-VHI}).
The same model was adopted for the U nucleus with a half-density radius of 6.8054~fm and
diffuseness parameters of 0.605~fm for protons and 0.786~fm for neutrons. 
The deformation of the U nucleus, which interacts with random
spatial orientation, was not taken into account.


\section{Analysis method}

\begin{figure}[hb!]
\centering
\resizebox{0.65\textwidth}{!}{%
\includegraphics*{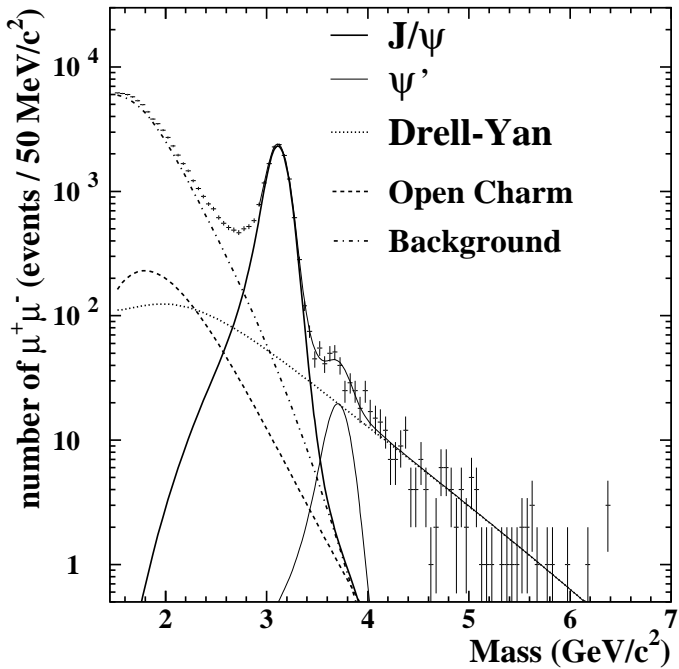}}
\caption{Opposite-sign dimuon invariant mass spectrum for the centrality class  
$35 < E_{\rm T} < 45$~GeV. The results from the fit are also shown.}
\label{fig:massfit}
\end{figure}

Particle yields have been extracted by fitting the dimuon mass spectrum, for each centrality bin.
First, a fit is performed to the $\mu^+\mu^+$ and $\mu^-\mu^-$ mass spectra, to determine 
the combinatorial background, mostly due to $\pi$ and $K$ decays.
The background in the opposite-sign dimuon sample is then parameterized according to the relation 
$N_{BG} = 2 \cdot R \cdot \sqrt{N^{++}\cdot N^{--}}$,
where $R$ can be taken as unity due to the absence of significant charge correlations
in Pb-Pb collisions. The validity of this relation is granted by a fiducial cut which rejects 
muon pairs where any of the muons would not have been accepted if
it had the opposite charge.

A multi-step fit to the $\mu^+\mu^-$ mass spectra (Fig.~\ref{fig:massfit} shows an example) 
is then performed to extract the four signal contributions which are relevant for masses above 2.5~GeV/$c^2$, 
namely \jpsi, \psip, Drell-Yan and open charm.
Finally, the acceptance corrections are  applied, thus obtaining the cross-sections ratio
$B_{\mu\mu}\sigma({\rm J}/\psi)/\sigma(DY)$.
This ratio of two dimuon processes has the advantage of being
insensitive to absolute normalizations (integrated luminosity) and to
most of the experimental efficiencies, allowing us to directly compare
these values to the normal nuclear absorption pattern derived from lighter
collision systems.
Furthermore, the Drell-Yan process is known (see e.g.\ Fig.~1 in Ref.~\cite{Abr97b} 
and Fig.~3 in Ref.~\cite{Abr03})
to scale linearly with the product $A \times B$ 
of the projectile and target mass numbers, and therefore linearly with the number of binary
nucleon-nucleon collisions.

The analysis is performed in the dimuon kinematical domain \mbox{$0<y_{\rm cms}<1$}
and \mbox{$-0.5 < \cos(\theta_{\rm CS}) < 0.5$}, where $y_{\rm cms}$ is the c.m.s.\ rapidity and
$\theta_{\rm CS}$ is the polar decay angle of the muons in the Collins-Soper reference 
frame~\cite{Col77}.  

For each of the four signal sources, muon pairs are generated and
propagated through the NA50 detector using the NA38/NA50 Monte Carlo simulation  
package which implements the multiple scattering and energy loss processes of    
muons as done in GEANT 3.21 (see e.g.\ \cite{Lou95}).                           
These events are then reconstructed using the standard NA50
reconstruction code in order to obtain the functional forms of each
signal, needed for the fits to the measured
invariant mass distributions, and their acceptances, needed for the calculation of the
cross-sections ratio.
For \jpsi\ and \psip\ generation, we use a gaussian rapidity distribution with an r.m.s.\ value
$\sigma = 0.6$ units~\cite{Lou95} and a uniform $\cos(\theta_{\rm CS})$ distribution. 
The transverse momentum is generated according to $K_1(M_{\rm T}/T)$, where $K_1$ is the modified Bessel 
function and $T=236$~MeV~\cite{Lou95}.
The Drell-Yan and open charm ($D\overline{D}$) contributions are calculated
with PYTHIA~6.124~\cite{PYTHIA} using the GRV~94~LO~\cite{GRV94} set of PDFs, as provided by the 
PDFLIB package~\cite{PDFLIB}.
In order to reproduce the measured $p_{\rm T}$ distributions,  
the width of PYTHIA's gaussian primordial $k_{\rm T}$ distribution has been 
set to 0.8~GeV/$c$ for Drell-Yan and 1.0~GeV/$c$ for open charm~\cite{Cap01}. 

The simulated and reconstructed dimuon mass distributions are parameterized by empirical functional 
forms. Such functions are illustrated in Fig.~\ref{fig:massfit}, after
adjustment to the measured invariant 
mass spectrum.
A first approximation of the \jpsi\ and \psip\ functional forms is
obtained through the Monte Carlo
procedure outlined above, which is meant to describe 
the detector momentum resolution, and the multiple
scattering and energy loss in the materials (target and muon filter)
crossed by the muons.
However, the very high statistical precision of our data
is incompatible with the imperfections of this Monte Carlo
description.  The calculated line shape of
the \jpsi\ must be slightly adjusted in order to closely reproduce the 
measured \jpsi\ peak~\cite{Ale04,Sha01}.

The line shape of the measured \jpsi\ resonance has been recently
studied in detail~\cite{Sha01}, resulting in the use of a functional
form which significantly improves the fit of the dimuon mass
distributions.  The new description of the \jpsi\ shape leads to a
decrease in the level of the \psidy\ cross-sections ratio by
about 6\,\% with respect to our preliminary results~\cite{Ram03},
while the pattern as a function of centrality remains unchanged (for
further details, see Refs.~\cite{Cas03} and~\cite{Sig03}).

The fit to the invariant mass distributions is made, using the maximum
likelihood method, through a multi-step procedure. First, the combinatorial
background in the $\mu^+\mu^-$ sample is estimated through a fit to the like-sign 
invariant mass spectra; then a fit in the region $2.9 < M < 8.0\ {\rm GeV}/c^2$ is performed
to obtain a first approximation of the \jpsi, \psip\ and Drell-Yan contributions;
then the open charm contribution is estimated by a fit in the mass region 
$1.7 < M < 2.6\ {\rm GeV}/c^2$.    
Finally, in order to minimize potential fit-induced distortions of the highly
sensitive Drell-Yan and \psip\ components of the mass spectrum, 
the last fit is performed again in the region $2.9 < M < 8.0\ {\rm GeV}/c^2$ (see
Fig.~\ref{fig:massfit}) with five free parameters, which are the
\jpsi, \psip\ and Drell-Yan
normalizations, the \jpsi\ mass $M_{\psi}$ and the \jpsi\ experimental width
parameter $\sigma_{\psi}$. 
The fitted values 
are in the range 97--100 MeV/$c^2$ for $\sigma_{\psi}$, 
and in the range 3.110--3.112 GeV/$c^2$ for $M_{\psi}$.

The acceptances of the NA50 detector, as calculated by our Monte Carlo
simulations and in the phase space window mentioned above, are given
in Table~\ref{tab:accept}, for \jpsi\ and Drell-Yan dimuons (for the
two mass regions considered in the present analysis).
The relative systematic error on the acceptances is estimated to be 2\,\%, while 
the statistical error is negligible.                                              

\begin{table}[htb]
\centering
\begin{tabular}{|c|c|} 
\hline
Process & Acceptance (\%) \\
\hline
\jpsi & 12.5 \\ 
Drell-Yan (2.9--4.5 GeV/$c^2$) & 13.8 \\
Drell-Yan (4.2--7.0 GeV/$c^2$) & 17.8 \\
\hline 
\end{tabular}
\caption{Acceptances of the experimental setup used in year
2000, for \jpsi~$\rightarrow \mu^+\mu^-$ and Drell-Yan dimuons, 
in the phase space window analyzed in this study.}
\label{tab:accept}
\end{table}


\section{Results}

\subsection{Summary of p-A results}

In order to accurately establish the normal nuclear absorption pattern, 
data were collected, with the NA50 apparatus, using the CERN SPS 400 or 450~GeV proton beam
and several different nuclear targets. Results from the first two of our three data samples
have been recently published~\cite{Abr03,Ale04}. A publication concerning the results from the third
data sample, taken at 400~GeV beam energy, is presently under preparation
(see Ref.~\cite{Bor04} for preliminary results). 

In a previous publication~\cite{Abr03} we presented the analyses of
several data sets for the (\jpsi)\,/\,Drell-Yan
cross-sections ratio.  That study merged measurements done with 450~GeV
protons (pp and pd from NA51; first 450 GeV \mbox{p-A} data samples from NA50), in the phase 
space domain
\mbox{$-0.5<\cos\theta_{\rm CS}<0.5$} and \mbox{$-0.4<y_{\rm
cms}<0.6$}, with S-U measurements done at 200~GeV/nucleon (from NA38),
in the domain \mbox{$-0.5<\cos\theta_{\rm CS}<0.5$} and
\mbox{$0.0<y_{\rm cms}<1.0$}.  For consistency reasons, we have
revisited that study and updated its results, including now the three \mbox{p-A} data
samples of NA50, to establish the normal nuclear absorption curve needed for the present work.  
Moreover, S-U results are not included anymore in this study.              
It is very interesting to check if the measured S-U \psidy\ results 
will become significantly different from the expected behaviour deduced   
using exclusively \mbox{p-A} measurements (see next subsection).                       
To bring the analysis of those \mbox{p-A} data sets (and also of the S-U data 
set) in tune with the procedures used in the present study of the Pb-Pb data,
the \jpsi\ and \psip\ resonances, in the fits of the mass
spectra, were described with the new line shapes~\cite{Ale04,Sha01}
and we consistently used the GRV~94~LO PDF sets in all calculations.
Moreover, the Monte Carlo si\-mu\-la\-tion of \jpsi\ production in \mbox{p-A} 
collisions has been 
upgraded in order to reproduce more accurately the measured \pt\ and 
$y$ distributions and thus obtain more precise \jpsi\ acceptance values for each
individual data sample (for more details, see Ref.~\cite{NA50-VHI}).

The cross-sections ratios \psidy\ for \mbox{p-A} collisions (including pp and pd  
results from NA51) are then fitted to a Glauber description of the normal        
nuclear absorption. The Glauber calculation is performed with three free         
parameters: the two independent normalizations (accounting for the                
different energy and kinematical conditions) 
and a common value for the \jpsi\ absorption cross-section.
The fit leads to $\sigma_{\rm abs}=4.18 \pm 0.35$~mb, a value very similar to our
previous result, $4.4 \pm 0.5$~mb (see Ref.~\cite{Abr03}). 
Energy, rapidity domain and isospin corrections are then applied 
to scale down the normalization of the absorption curve 
from a pp system at 450 GeV to a Pb-Pb system at 158~GeV.           
To perform this scaling, we first compare the \jpsi\ absolute \mbox{p-A}    
cross-sections measured by NA50 at 450 and 400 GeV with the \jpsi\  
absolute \mbox{p-A} cross-sections measured by NA38 and NA3~\cite{NA3}     
at 200 GeV, in the phase space domain                               
\mbox{$-0.5 < \cos(\theta_{\rm CS}) < 0.5$} and                     
\mbox{$0.0 < y_{\rm cms} < 1.0$}.                                  
If these absolute cross-sections are fitted with a common          
$\sigma_{\rm abs}$, 
we obtain a value of $4.11 \pm 0.43$~mb, in excellent agreement with the
value of $4.18 \pm 0.35$~mb quoted above and measured using
\psidy\ ratios at 450 and 400~GeV\footnote{In the following,
we adopt the value $4.18 \pm 0.35$~mb which is insensitive to most of the
systematic uncertainties and, therefore, can be considered as our best
estimate of $\sigma_{\rm abs}$.}.  
The same fit leads to a factor of $0.319 \pm 0.025$ 
($0.343 \pm 0.027$) which allows to scale down
the \jpsi\ absolute cross-section from the 450 (400) GeV  
kinematical domain to the 200 GeV kinematical domain. In addition,  
an energy correction factor of $0.737 \pm 0.006$ is used to scale down
the \jpsi\ cross-section from 200 to 158 GeV.                      
This correction was obtained using the parametrization 
$\sigma_{\psi}(\sqrt{s}) = \sigma_0 ( 1 - M_{\psi}/\sqrt{s} )^n$, with
the value $n = 12.8 \pm 0.3$ derived from a fit to available
measurements.
Furthermore, the change induced by the different $\sqrt{s}$ on the 
$x_F$ window (for a fixed $y_{\rm cms}$ window: $0.0 < y_{\rm cms} < 1.0$) 
and on the $x_F$ distribution (parametrized as in Ref.~\cite{Sch94}) 
implies a very small correction: $1.020 \pm 0.013$.             
The corresponding scaling factors for the Drell-Yan cross-section are 
computed at leading order using the GRV~94~LO PDF sets.        
The simultaneous change in energy and kinematical domains from 
450 (400) GeV to 200 GeV amounts to $0.504 \pm 0.012$           
($0.544 \pm 0.010$), while the change in energy and $x_F$ for   
the much smaller drop between 200 and 158~GeV gives 0.7085 and  
1.091, respectively. Finally, a factor 0.969 is applied to the  
absorption curve corresponding to the isospin change            
from pp to Pb-Pb.                                               

The statistical uncertainty on the absorption curve as measured at the higher energies  
is $\pm 1$\,\% at low centrality reaching $\pm 4$\,\% at high centrality.               
Adding in quadrature the errors on the rescaling factors we obtain at 158 GeV an       
uncertainty of $\pm 8.3$\,\% at low centrality reaching $\pm 9.0$\,\% at high centrality.
This normal nuclear absorption curve is reported in the following figures along with
the Pb-Pb \psidy\ data points.

\subsection{Pb-Pb results}

\begin{figure}[p]
\centering
\begin{tabular}{cc}
\resizebox{0.48\textwidth}{!}{\includegraphics*{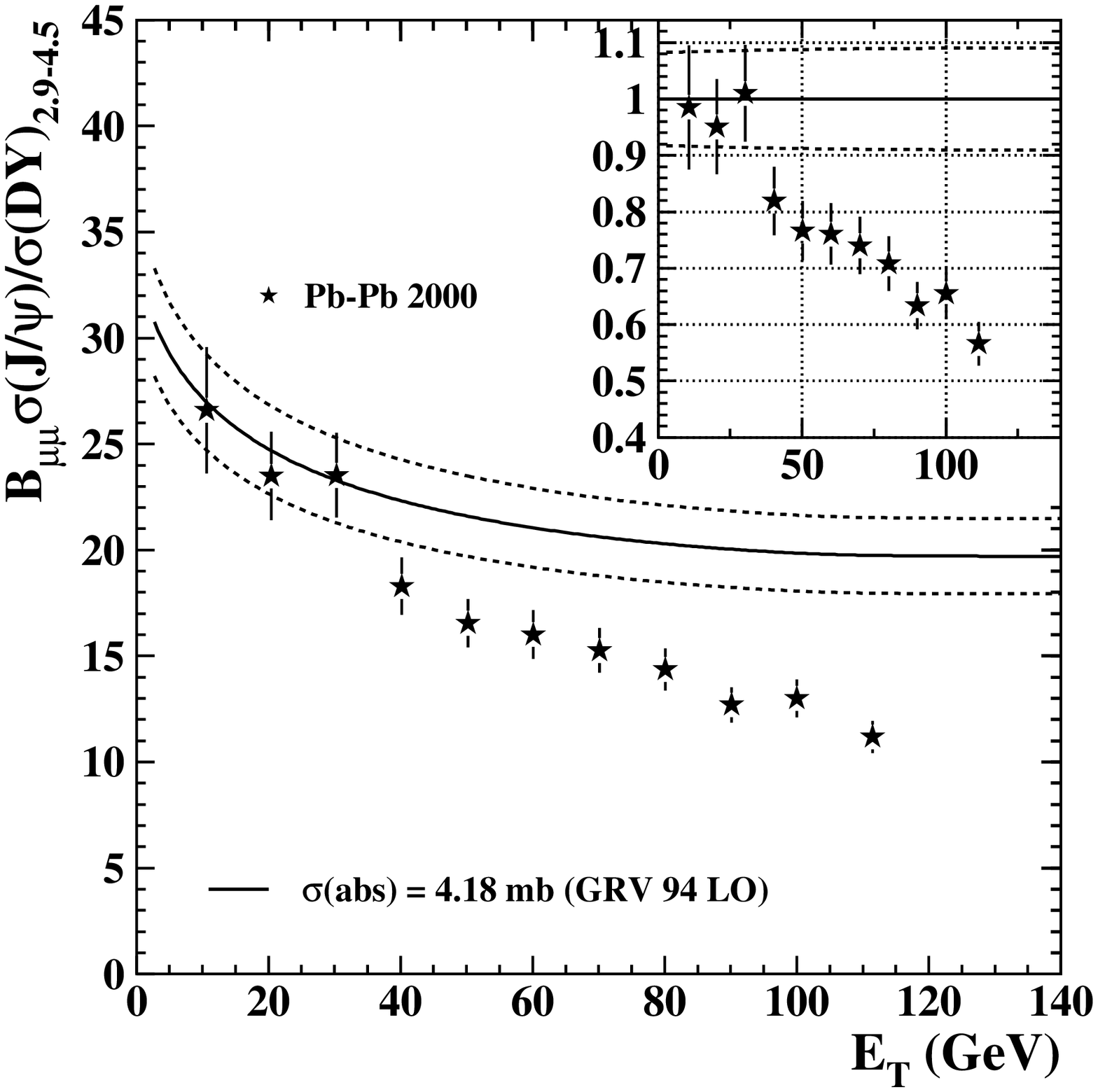}} &
\resizebox{0.48\textwidth}{!}{\includegraphics*{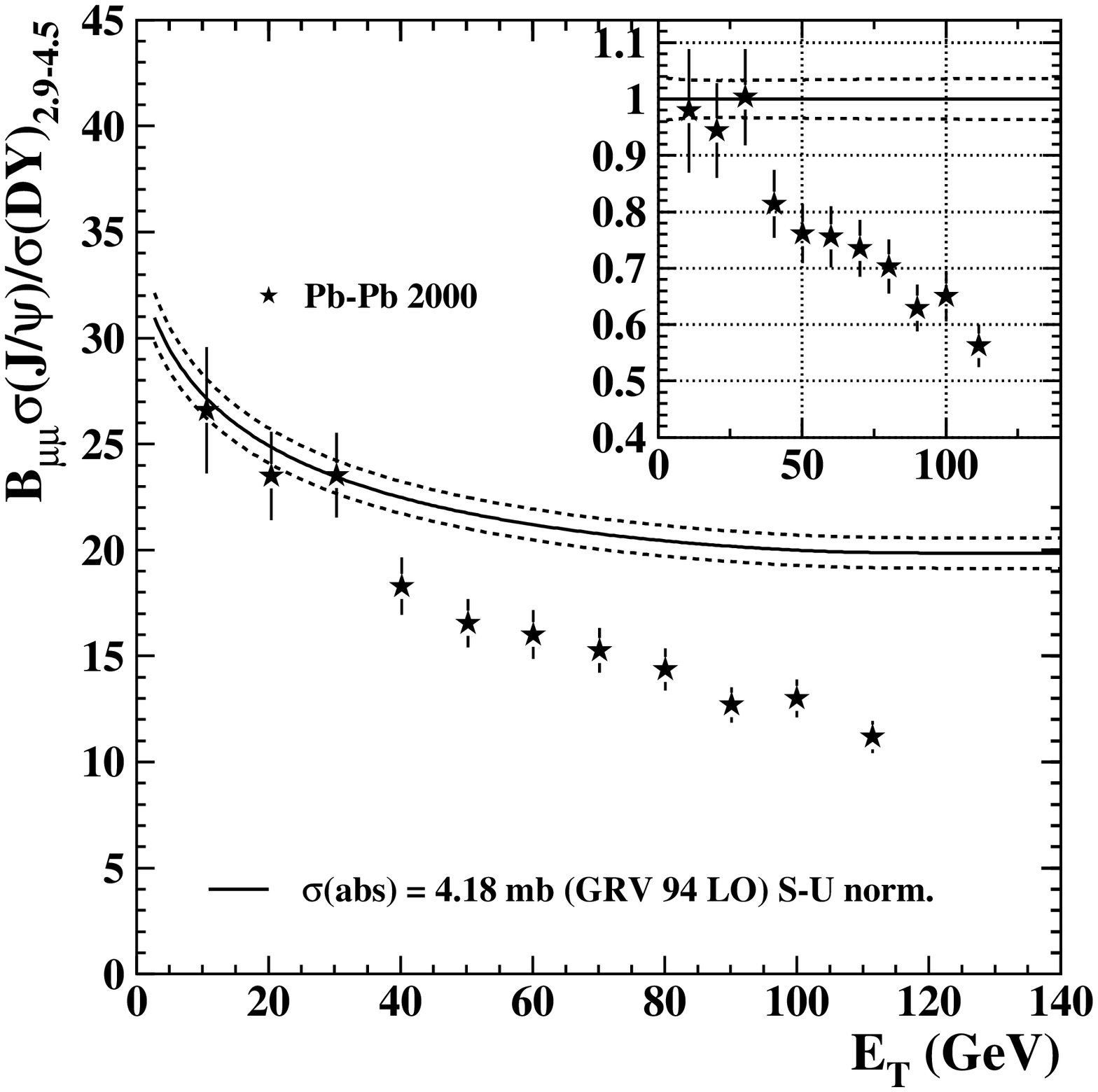}} \\
\end{tabular}
\caption{The \psidy\ cross-sections ratio as a function of transverse energy for the Pb-Pb 2000 data 
sample (left). The normal absorption curve is presented together with the combined error from the
Glauber fit and the rescaling procedure (dashed curves).
The inset shows the ratio Measured/Expected, i.e.\ data over normal nuclear absorption. 
The right panel presents the same data compared to the absorption curve computed using also S-U NA38
data for $\sigma_{\rm abs}$ determination and curve normalization.}
\label{fig:psiet00}
\vspace{2mm}
\centering
\resizebox{0.55\textwidth}{!}{%
\includegraphics*{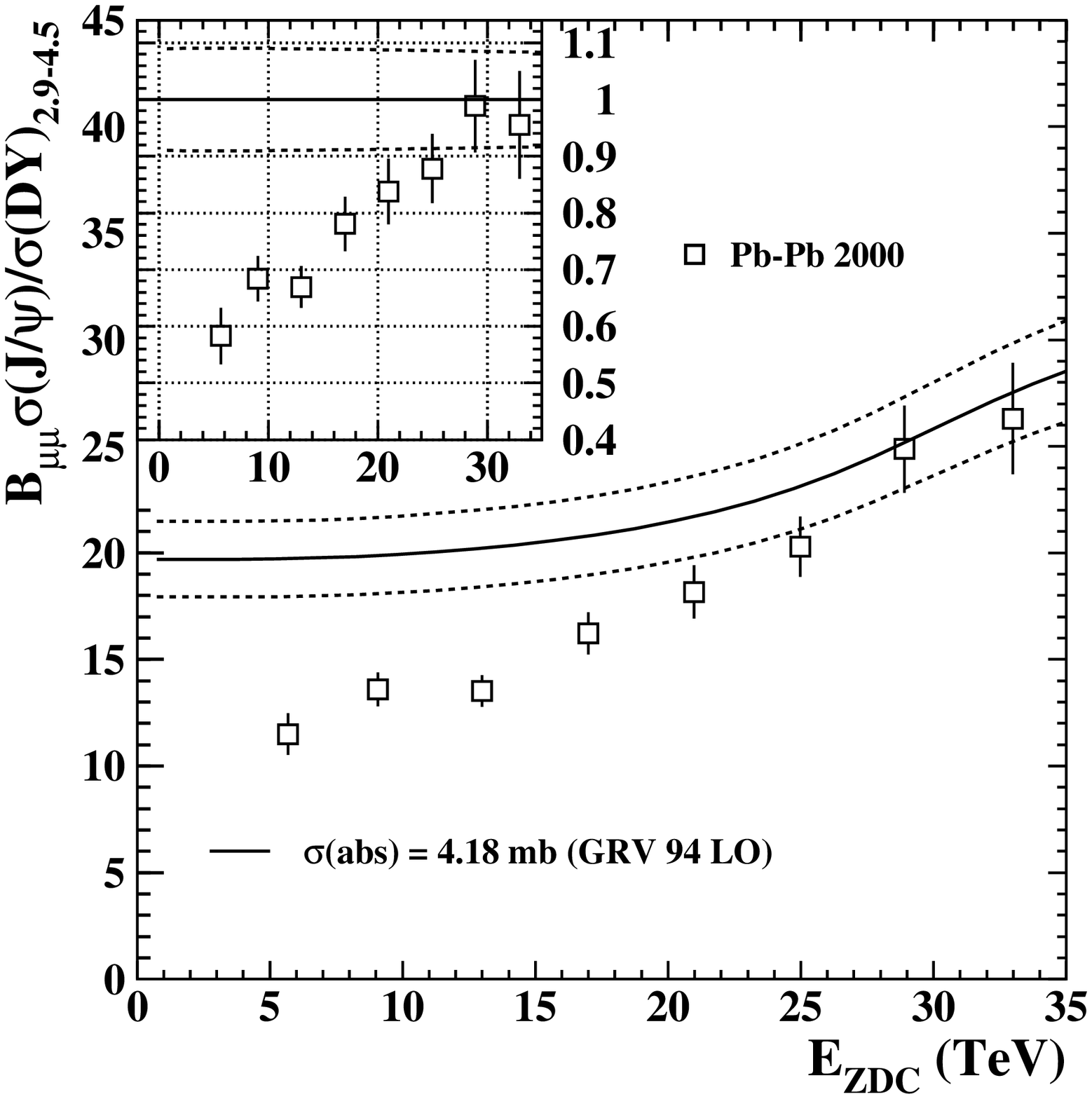}}
\caption{The \psidy\ cross-sections ratio as a function of forward energy for the Pb-Pb 2000 data sample.
The inset shows the ratio Measured/Expected, i.e.\ data over normal nuclear absorption.}
\label{fig:psizdc00}
\end{figure}

\begin{figure}[ht]
\centering
\resizebox{0.55\textwidth}{!}{\includegraphics*{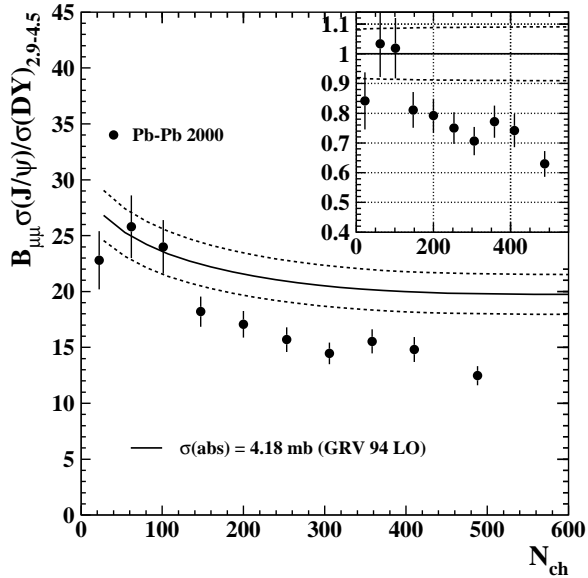}} 
\caption{The \psidy\ cross-sections ratio as a function of charged mul\-ti\-pli\-ci\-ty 
for the Pb-Pb 2000 data sample.
The inset shows the ratio Measured/Expected, i.e.\ data over normal nuclear absorption.}
\label{fig:psidnde00}
\end{figure}

We now present the 
\jpsi\ suppression pattern for Pb-Pb interactions, as obtained from the
data collected in year 2000. 
As in the past, we consider the cross-sections ratio
\psidy, where the Drell-Yan differential cross-section is integrated in the mass domain
2.9--4.5~GeV/$c^2$.  The results from three independent
analyses, using the centrality variables \et, \ezdc\ and \dnde, are
shown in Figs.~\ref{fig:psiet00}, \ref{fig:psizdc00}
and~\ref{fig:psidnde00}, which also include the normal nuclear absorption
curve, obtained as explained above.  The corresponding numerical
values are reported in Table~\ref{tab:psidy}. The results referred to the high mass Drell-Yan
domain (4.2--7.0~GeV/$c^2$), which are less sensitive to the specific
PDF sets used in the analysis,
can be obtained by scaling up those reported in Table~\ref{tab:psidy}
with the factor 7.96, the ratio between the Drell-Yan cross-sections
integrated in the two mass domains.

\begin{table}[htb]
\centering
\begin{tabular}{|cc|cc|cc|} 
\hline
Class & B$_{\mu\mu}\sigma$(J/$\psi$)\,/\,$\sigma$(DY) 
& Class & B$_{\mu\mu}\sigma$(J/$\psi$)\,/\,$\sigma$(DY)  
& Class &  B$_{\mu\mu}\sigma$(J/$\psi$)\,/\,$\sigma$(DY) \\
\hline
$1_T$ &$26.6 \pm3.0$& $1_F$ &$26.3 \pm2.6$&$1_N$ &$22.8 \pm2.6$\\ 
$2_T$ &$23.5 \pm2.1$& $2_F$ &$24.9 \pm2.1$&$2_N$ &$25.8 \pm2.8$\\ 
$3_T$ &$23.5 \pm2.0$& $3_F$ &$20.3 \pm1.4$&$3_N$ &$24.0 \pm2.4$\\            
$4_T$ &$18.3 \pm1.3$& $4_F$ &$18.2 \pm1.2$&$4_N$ &$18.2 \pm1.4$\\           
$5_T$ &$16.5 \pm1.1$& $5_F$ &$16.2 \pm1.0$&$5_N$ &$17.1 \pm1.2$\\         
$6_T$ &$16.0 \pm1.1$& $6_F$ &$13.5 \pm0.7$&$6_N$ &$15.7 \pm1.1$\\     
$7_T$ &$15.3 \pm1.0$& $7_F$ &$13.6 \pm0.8$&$7_N$ &$14.5 \pm1.0$\\        
$8_T$ &$14.4 \pm1.0$& $8_F$ &$11.5 \pm1.0$&$8_N$ &$15.5 \pm1.1$\\
$9_T$ &$12.7 \pm0.8$&       &             &$9_N$ &$14.8 \pm1.1$\\          
$10_T$&$13.0 \pm0.9$&       &             &$10_N$&$12.5 \pm0.9$\\         
$11_T$&$11.2 \pm0.8$&       &             &      &             \\ 
\hline
\end{tabular}
\caption{Cross-sections ratio \psidy, referred to the 2.9--4.5~GeV/$c^2$ Drell-Yan mass domain,
for the Pb-Pb 2000 data, with three independent centrality estimators. Values for the Drell-Yan mass
domain 4.2--7.0~GeV/$c^2$ may be obtained as explained in the text. 
Classes are defined in Tables~\ref{tab:centralet}, \ref{tab:centralezdc} and \ref{tab:centraldnde}. 
Only statistical errors are given; systematic errors are negligible in comparison to statistical
ones, as explained in the discussion session.}
\label{tab:psidy}
\end{table}

Figure~\ref{fig:psiet00} (left) shows that the \psidy\ ratio             
in peripheral Pb-Pb collisions is perfectly consistent with the pattern of normal nuclear 
absorption, as deduced from \mbox{p-A} collisions alone, with the appropriate  
normalization to 158~GeV/nucleon and to Pb-Pb isospin content.
The departure from the normal absorption pattern
at $E_{\rm T}\sim 35$~GeV and the non saturation at high \et,
already observed in previously published analyses,
is also seen in this new sample of Pb-Pb collisions.
The same observations can be made about the \jpsi\ absorption pattern as a function
of the second centrality variable, \ezdc, as shown in
Fig.~\ref{fig:psizdc00}. 

In Fig.~\ref{fig:psiet00} (right) we compare the Pb-Pb data to our previous determination
of the absorption curve which made use of the NA38 S-U \psidy\ ratios
at 200 GeV together with the best estimates of the p-A data (either using
\jpsi\ or \psidy\ results) at higher energies~\cite{Bor04}.
This previous determination led to $\sigma_{\rm abs} = 4.18 \pm 0.35$~mb.
It is truly remarkable that this is exactly identical to the value
presented above, and which was obtained through a rather different
procedure, exclusively based on the 400 and 450~GeV proton-nucleus data.
However, while the two determinations of $\sigma_{\rm abs}$ coincide,
the new method leads to an absorption curve with a significantly larger
error band (as seen in Fig.\ref{fig:psiet00} left) originating from the uncertainty on
the energy-rapidity rescaling factors, which are no longer constrained
by the 200 GeV S-U results.

\begin{figure}[p]
\centering
\begin{tabular}{cc}
\resizebox{0.48\textwidth}{!}{\includegraphics*{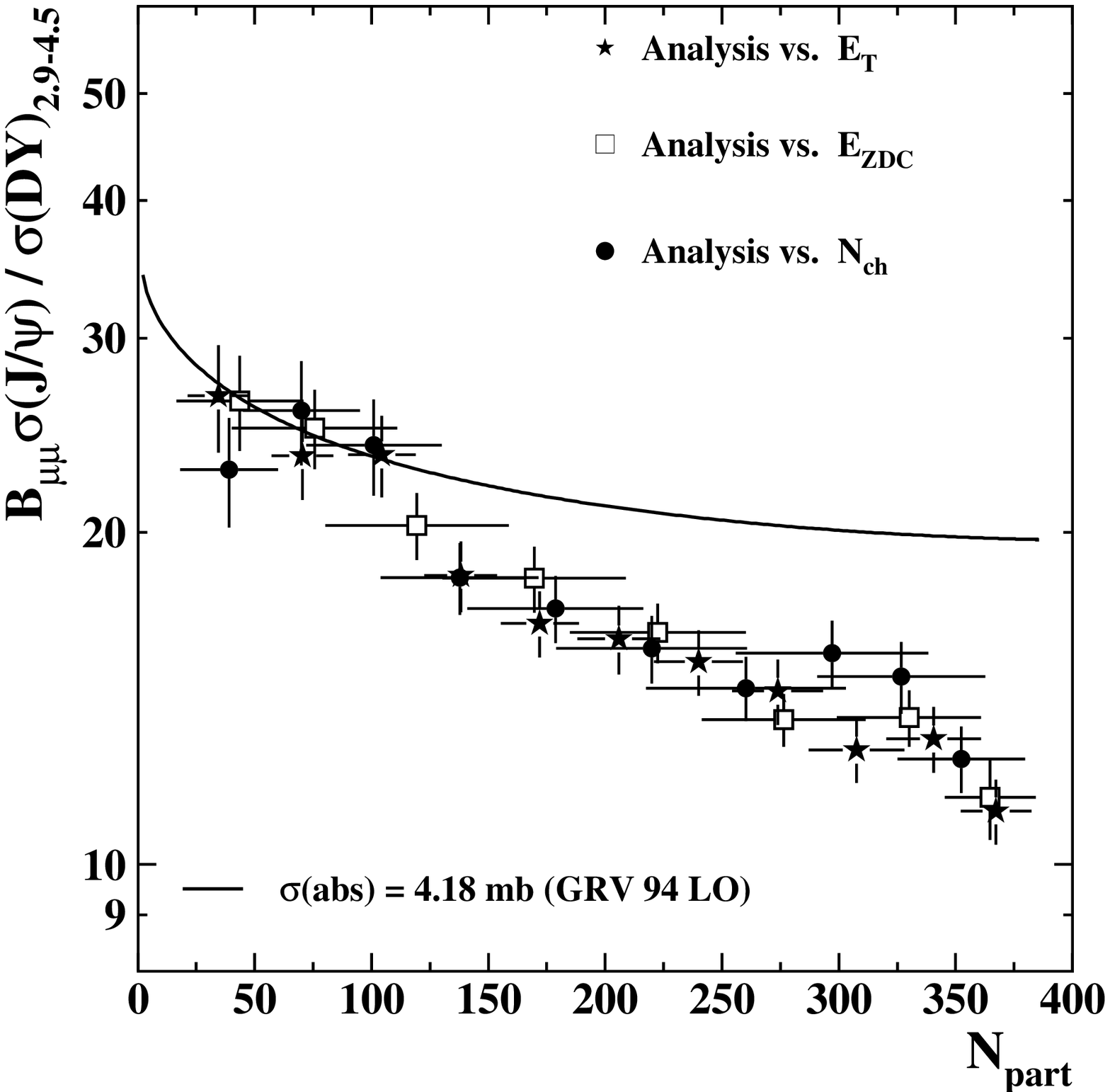}} &
\resizebox{0.48\textwidth}{!}{\includegraphics*{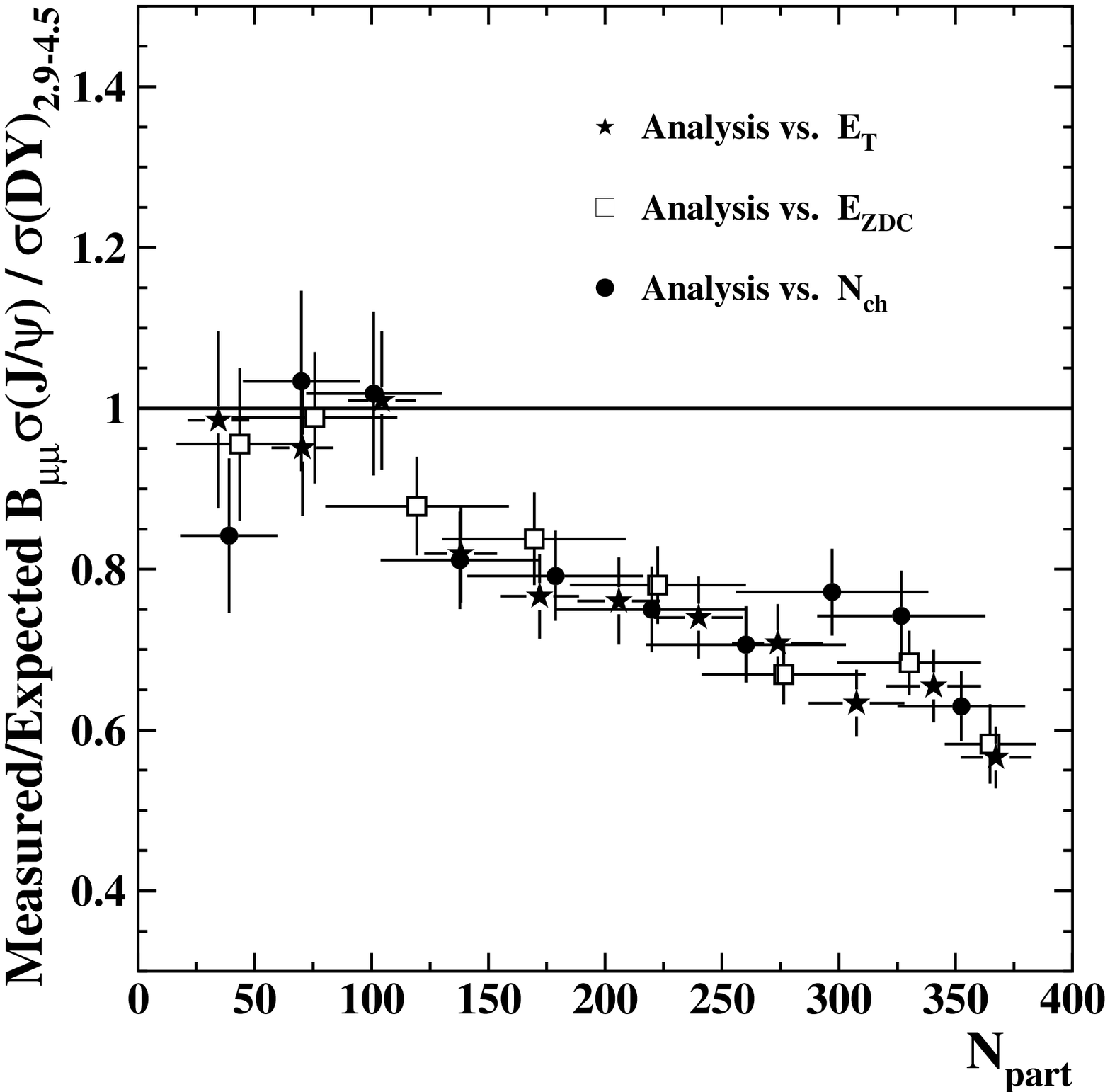}} \\
\end{tabular}
\caption{The \psidy\ cross-sections ratio as a function of $N_{\rm part}$ for three analyses of the 
Pb-Pb 2000 data sample, compared to (left) and divided by (right) the
normal nuclear absorption pattern.}
\label{fig:psidyn}
\vglue2mm
\centering
\begin{tabular}{cc}
\resizebox{0.48\textwidth}{!}{\includegraphics*{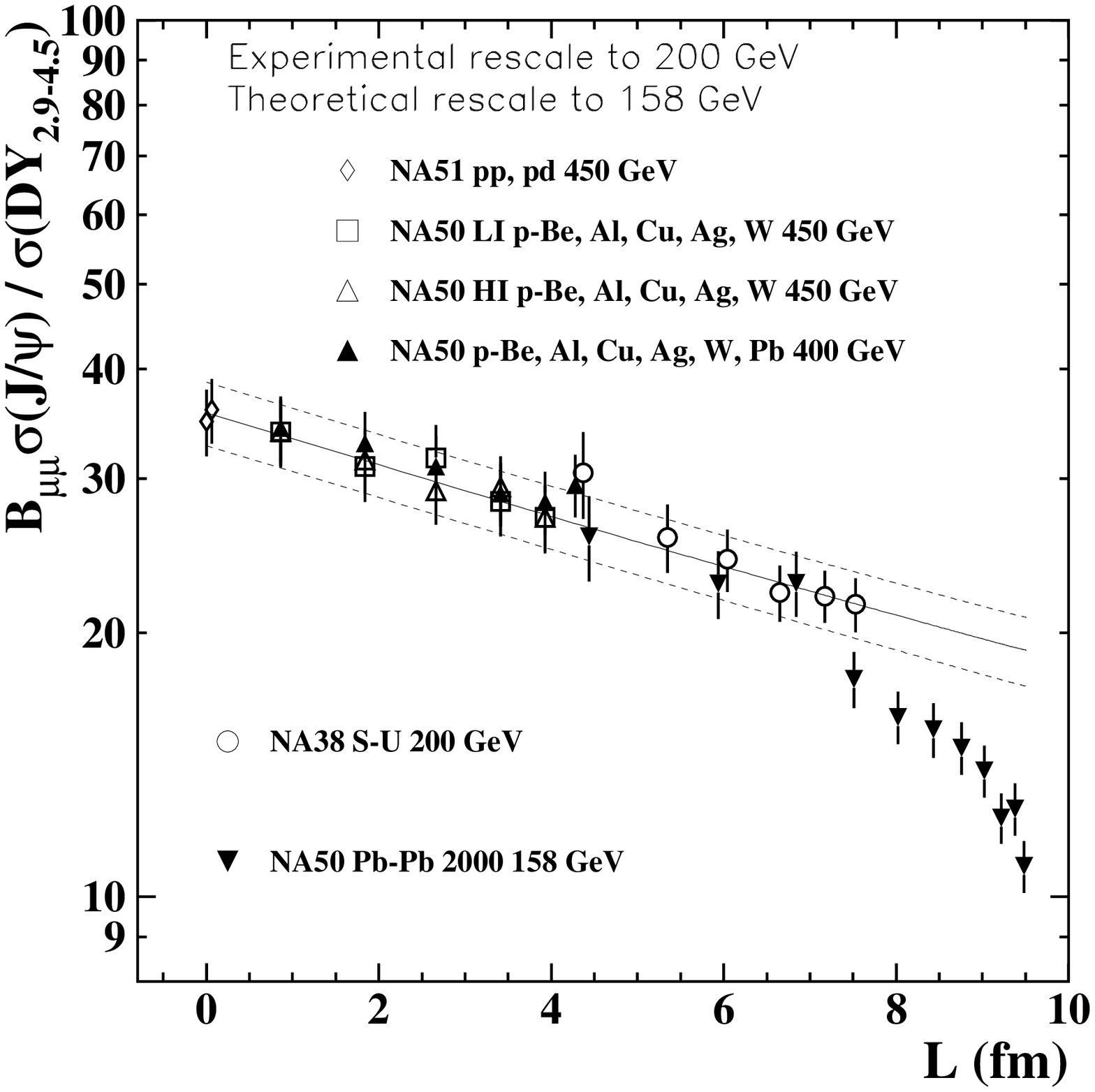}} &
\resizebox{0.48\textwidth}{!}{\includegraphics*{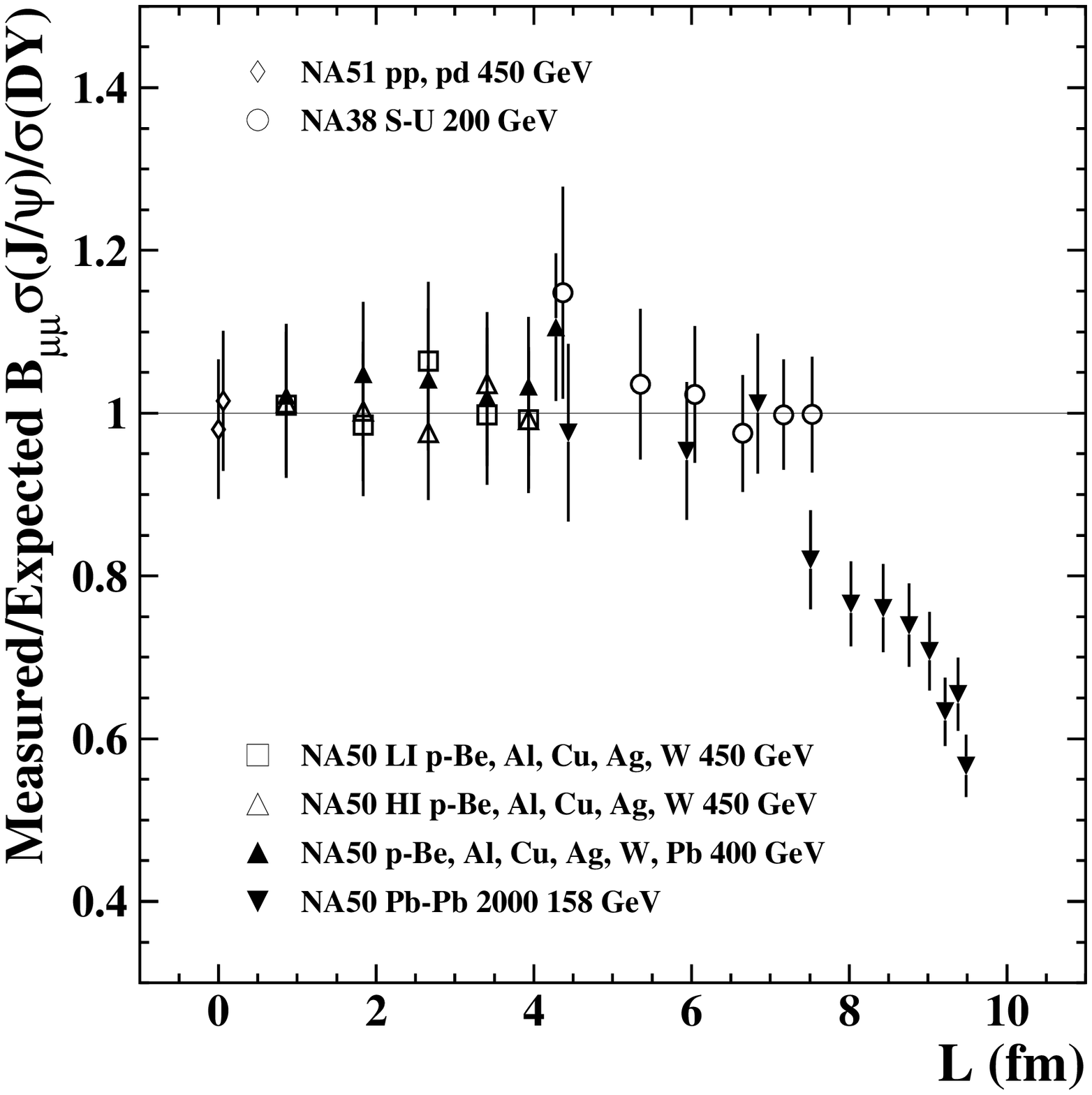}} \\
\end{tabular}
\caption{The \psidy\ cross-sections ratio vs.\ $L$, for several
  collision systems, compared to (left) and divided by (right) the
  normal nuclear absorption pattern.  The measured data have been
  rescaled to 158~GeV/nucleon and for proton-proton isospin. 
  The \et\ data set was used for the Pb-Pb points.}
\label{fig:ratvsl}
\end{figure}

The analysis as a function of the third centrality variable, \dnde,
has the unique feature
that the charged particle multiplicity is evaluated in the
rapidity window where the dimuons are measured by the muon spectrometer.
From the \psidy\ results displayed in Fig.~\ref{fig:psidnde00} we observe that the suppression
pattern is very similar to the one obtained with the \et\ centrality variable.
Note that the most peripheral data point corresponds to about half the 
statistics of the semi-central and central data points.
                                     
A comparison of the three independent analyses is presented in
Fig.~\ref{fig:psidyn}, using the average 
number of participant nucleons, $N_{\rm part}$,
as a common centrality variable.
It is worth recalling that \et\ and the charged particle rapidity
density at mid-rapidity are directly related to the energy density
reached in the collision, through the Bjorken formula, while the
forward energy measured in the ZDC is more strongly correlated to the
geometry of the collision, being a simple and robust estimator of
$N_{\rm part}$.
We have chosen \npart\ for the purpose of comparing our three
independent analyses because it is well known that, at SPS energies,
both \et\ and \dnde\ are linearly proportional to \npart, as expected
in the framework of the wounded nucleon model, up to the most central
collisions (see, for instance, Refs.~\cite{Kha97} and~\cite{DNDETA2}).

The horizontal error bars in Fig.~\ref{fig:psidyn}
represent the r.m.s.\ values of the \npart\ distribution in that bin,
which depend on 
the experimental smearing, specific of each centrality variable. 
We see from this figure that the three (completely independent)
centrality measurements give a very consistent picture.  In
particular, we do not see any evidence that the \jpsi\ suppression
pattern looks different when looked as a function of variables related
to particle production (\et\ and \dnde) or as a function of simple
geometry (\ezdc).

In order to compare our Pb-Pb results with other (lighter) collision systems, 
we present in Fig.~\ref{fig:ratvsl} (left) the \psidy\ cross-sections ratio obtained
in this analysis together with the results obtained in S-U (NA38) and \mbox{p-A} (NA50)
interactions.  
The right panel of this figure shows the same data points after dividing 
by the function representing the normal nuclear absorption, using for 
the horizontal scale the average length $L$ of nuclear matter
traversed by the $c\bar{c}$ state. The length $L$ has been evaluated
as $\langle\rho(r) L\rangle/\rho_0$, where $\rho(r)$ is the nuclear density, normalized for each
nucleus in order to give $A$ nucleons upon integration, and $\rho_0$ is the average
nuclear density, $0.17$~fm$^{-3}$. 
The figure shows how the \jpsi\ anomalous suppression 
clearly emerges from the reference line obtained exclusively from proton-nucleus data.
We observe also that the S-U data points                                            
closely match the absorption curve determined from the \mbox{p-A} data,           
leaving very little room for anomalous absorption of the \jpsi\ in S-U collisions.      

\begin{figure}[ht]
\centering
\begin{tabular}{cc}
\resizebox{0.48\textwidth}{!}{\includegraphics*{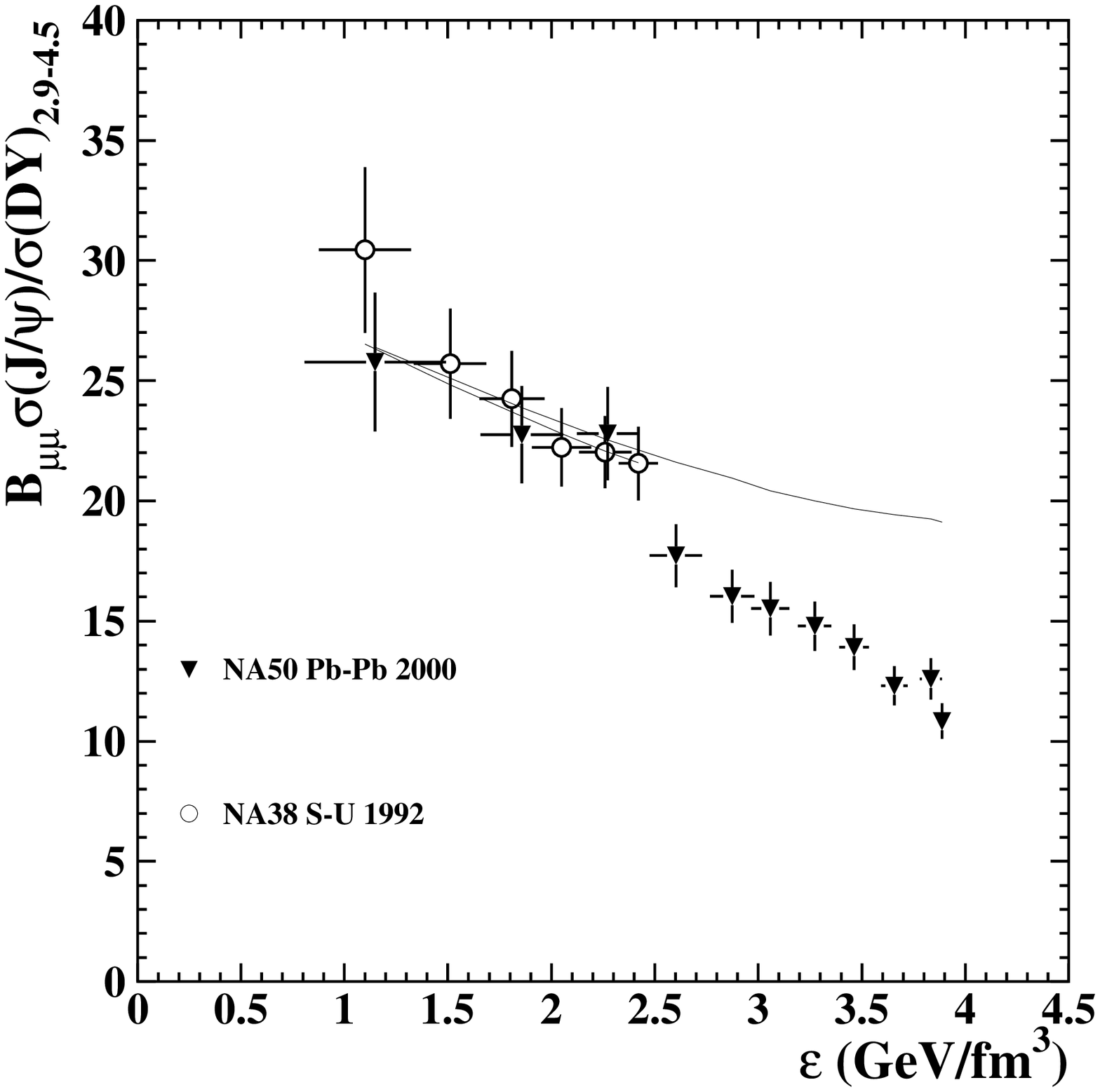}}&
\resizebox{0.48\textwidth}{!}{\includegraphics*{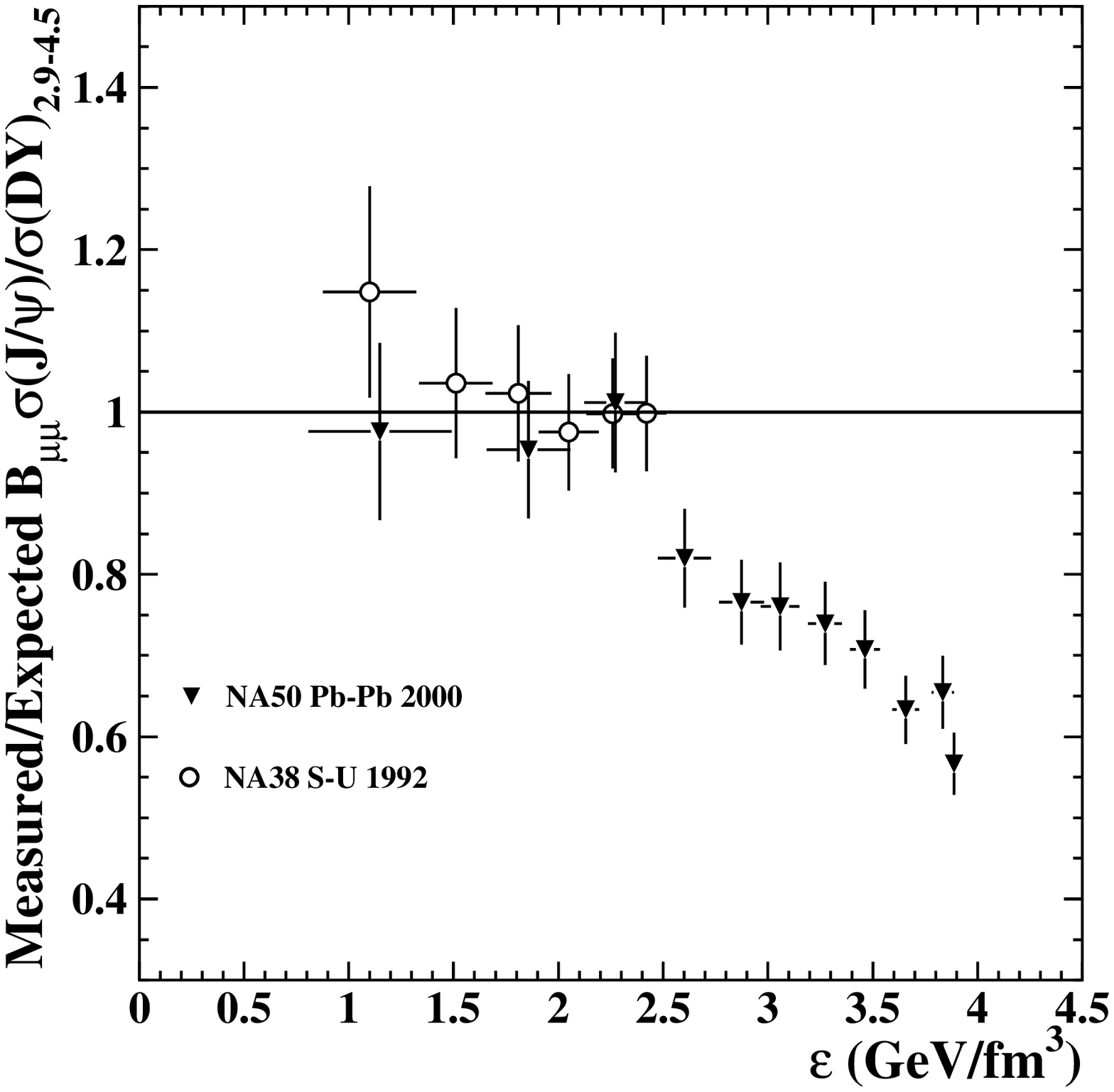}} \\
\end{tabular}
\caption{The \psidy\ cross-sections ratio as a function of the energy
density, for the S-U and Pb-Pb collision systems, compared to (left) and divided by (right)
the normal nuclear absorption pattern. 
The measured data have been rescaled to 158~GeV/nucleon and for proton-proton isospin. 
The \et\ data set was used for the Pb-Pb points.}
\label{fig:ratvsepsil}
\end{figure}

Finally, we present in Fig.~\ref{fig:ratvsepsil} the \psidy\ cross-sections ratio 
obtained in S-U (NA38) and Pb-Pb (NA50) interactions as a function of $\epsilon$,
the energy density averaged over the whole transverse area of the collision, 
evaluated with the Bjorken formula:
$$\epsilon={\dd E_{\rm T}/\dd\eta|_{\rm max} \over {c \tau A_{\rm T}}}\quad ,$$
where $A_{\rm T}$ is the overlap area in the transverse plane and 
$\tau$ is the formation time, assumed to be 1~fm/$c$.  We obtained the
total transverse energy scaling up the measured (neutral) \et\ by a
factor 3.
The different rapidity coverages of the NA38 and NA50 electromagnetic
calorimeters have been taken into account, knowing that the $\dd
E_{\rm T}/\dd\eta$ distributions depend on the collision centrality.

This figure shows that the departure from the normal nuclear absorption curve 
sets in for energy densities around 2.5~GeV/fm$^3$, just above the values             
reached in the most central S-U collisions.   
The absorption curves for S-U and Pb-Pb in the left panel
of Fig.~\ref{fig:ratvsepsil} are slightly different because the relation between
energy density and $L$ (obtained from the Glauber calculation) depends on the colliding nuclei.


\section{Discussion}

We discuss first the statistical and systematical errors of the Pb-Pb 2000 results.
Statistical errors in a given centrality bin, coming from the fit to the mass
spectrum, are in the range 5.5--8\,\%, and are, of course, dominated by the Drell-Yan
statistics.
We have evaluated the systematic errors coming from several sources
(for details see Refs.~\cite{Cas03,San04,Sig03}): 
(i)~variation
of the mass fit starting point, (ii)~variation of the level of the
muon target cut, 
(iii)~variation of the \et--\ezdc\
correlation cut from $\pm 3\sigma$ to $\pm 2\sigma$, 
(iv)~use of a counting technique using the number of signal events in
the mass ranges 2.9--3.3 and 4.2--7.0~GeV/$c^2$, with subtraction of the number of
Drell-Yan events in the first mass range.
The point-to-point systematic error from the above sources was found to be negligible
when compared to the statistical error.

Another systematic effect comes from the choice of PDFs for the Drell-Yan
process (influencing both the mass fit and the acceptance calculation),
which affects the overall normalization of our results while leaving
unchanged the shape of the suppression pattern as a function of centrality. 
We have compared our standard choice of PDF (GRV~94~LO) with more recent
PDFs computed at leading order, namely GRV~98~LO~\cite{GRV98}, 
MRST (central gluons)~LO~\cite{MRST98} and CTEQ~5L~\cite{CTEQ5L}.
We have studied the effect of changing the Drell-Yan functional form while keeping the
same normalization in the 4.2--7.0~GeV/$c^2$ mass region, as imposed by our data.
The maximum change in the Pb-Pb \psidy\ cross-sections ratio is 3.5\,\%
(for comparison, using the sets GRV~92~LO and MRS~A (Low Q$^2$) quoted in the  
introduction, we obtain a change of 0.8\,\% and 10\,\%, respectively).

We now discuss briefly how the Pb-Pb 2000 results on the \psidy\ cross-sections ratio,
presented above, compare with our results from previous Pb-Pb data
taking periods.

\begin{figure}[h!]
\centering
\begin{tabular}{cc}
\resizebox{0.48\textwidth}{!}{\includegraphics*{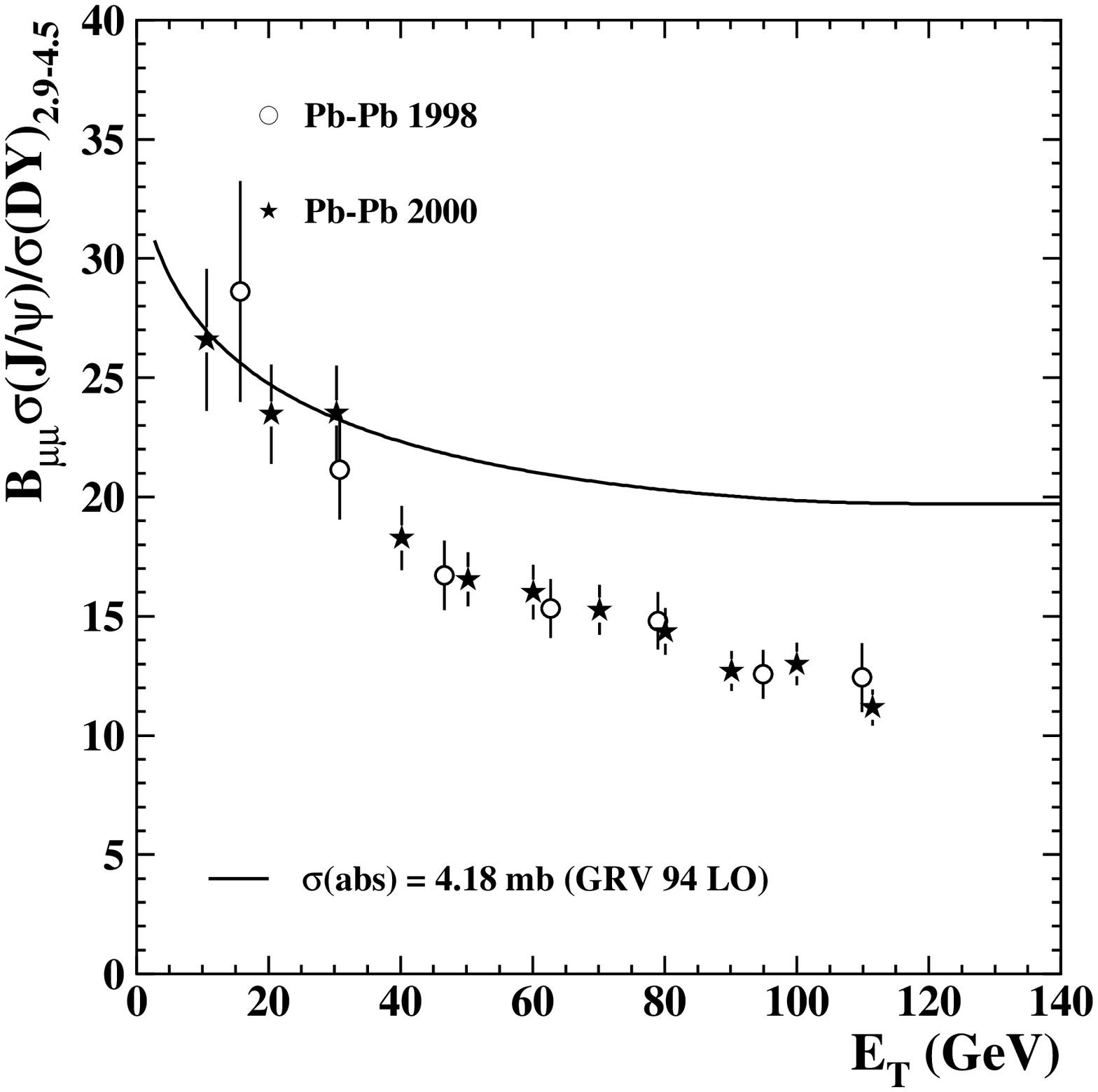}} &
\resizebox{0.48\textwidth}{!}{\includegraphics*{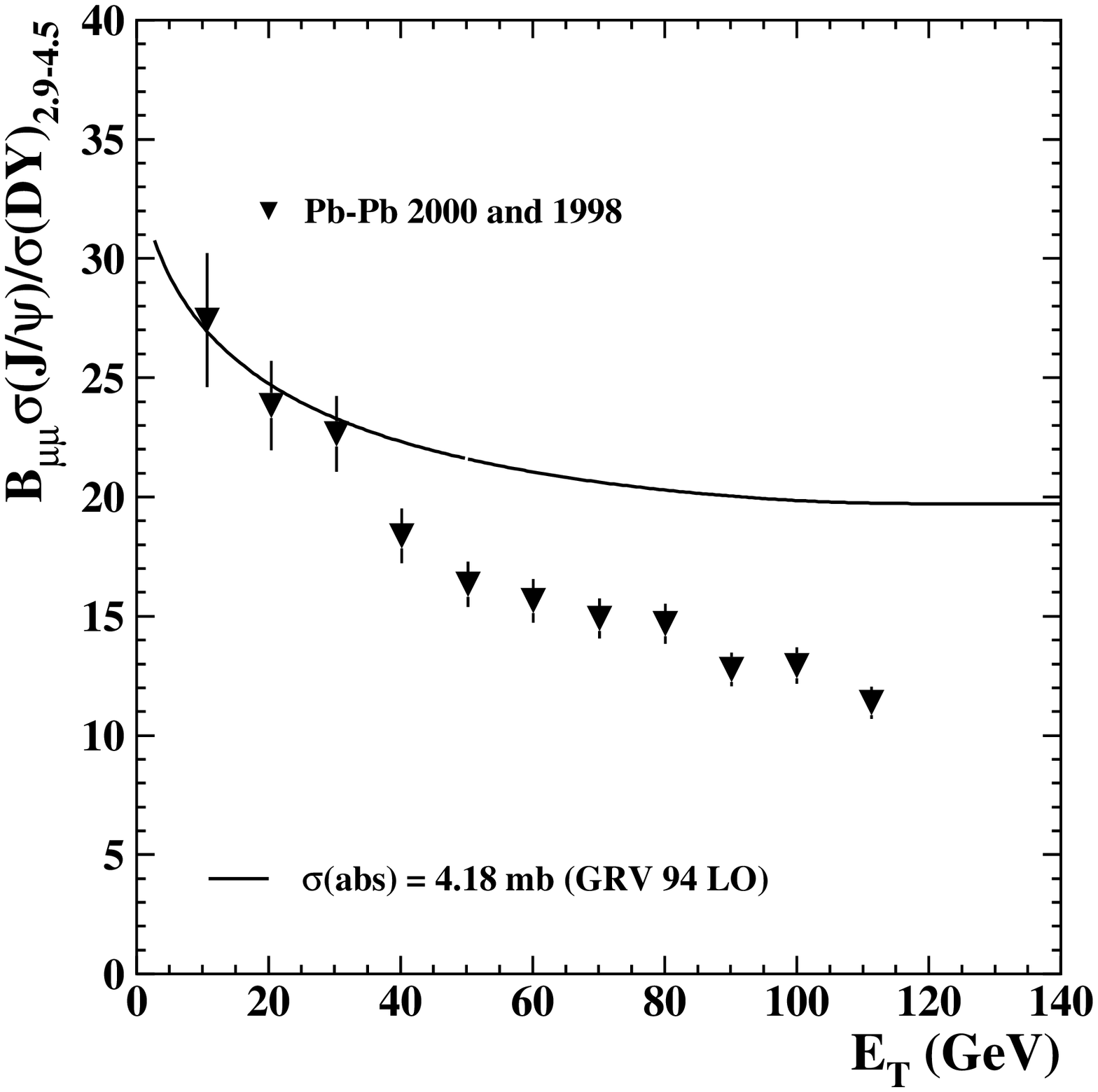}} \\
\end{tabular}
\caption{Comparison between 1998 and 2000 results: separate \psidy\ cross-sections ratios 
from 1998 and 2000 (left) and weighted average (right), as a function of 
transverse energy.}
\label{fig:comp9800}
\end{figure}

We consider the 1998 data sample, collected with an experimental setup similar to the one
used in year 2000 except for the vacuum around the target,
and find excellent agreement, as shown in the left panel of Fig.~\ref{fig:comp9800},
after repeating the analysis with our most recent procedures (see Ref.~\cite{San04}). 
The introduction of the vacuum around the target region has brought a considerable increase
in statistics of peripheral events. 
The weighted average of the two analyses (where data of year 1998 have been rebinned using the
year 2000 centrality classes) is presented in the right panel of Fig.~\ref{fig:comp9800}.
The resulting suppression pattern suggests two different suppression
regimes, one for peripheral and one for central reactions. The
observed transition between these two regimes, including the smearing
due to the resolution of the e.m. calorimeter which amounts here to 8\%,
extends over a range of about 15~GeV in \et.
     
\section{Conclusion}

We have analyzed the most recent NA50 Pb-Pb data sample, collected in year 2000 under
improved experimental conditions, which allowed us to extend the analysis
of the (\jpsi)\,/\,Drell-Yan cross sections ratio down to very peripheral interactions
using only dimuon triggers.
We have used three different centrality estimators, namely transverse energy, forward energy
and charged particle multiplicity.
Our analysis shows that peripheral Pb-Pb interactions, with impact parameter $b>8.5$~fm,
exhibit a \jpsi\ production yield
in agreement with the normal nuclear absorption pattern, with $\sigma_{\rm abs}=4.18$~mb,
derived from an extensive study of \mbox{p-A}                                   
collisions. For smaller impact parameter values we observe a departure from the normal 
absorption curve, followed by a persisting decrease up to the most central 
Pb-Pb collisions.  

\section*{Acknowledgements}

We would like to acknowledge the constant effort of the CERN PS, SPS and EA groups, and in 
particular of Lau Gatignon, for providing a high quality Pb beam to our experiment. 
We are grateful to our colleagues C. Baglin and A. Bussi\`ere of LAPP, Annecy (France)
for their contribution to the hardware and software of the readout system.
We thank Ramona Vogt for useful discussions concerning parton distribution functions.
This work was partially supported by the Funda\c{c}\~ao para a Ci\^encia e a Tecnologia, Portugal.


\begin{thebibliography}{99}
\baselineskip=10pt

\bibitem{Mat86} T. Matsui and H. Satz, Phys. Lett. B 178 (1986) 416. 
\bibitem{Abr97a} M.C. Abreu et al. (NA50 Collaboration), Phys. Lett. B 410 (1997) 327. 
\bibitem{Abr97b} M.C. Abreu et al. (NA50 Collaboration), Phys. Lett. B 410 (1997) 337. 
\bibitem{Abr98}  M.C. Abreu et al. (NA51 Collaboration), Phys. Lett. B 438 (1998) 35.   
\bibitem{Abr99a} M.C. Abreu et al. (NA38 Collaboration), Phys. Lett. B 449 (1999) 128.  
\bibitem{Abr99b} M.C. Abreu et al. (NA38 Collaboration), Phys. Lett. B 466 (1999) 408.  
\bibitem{Abr99} M.C. Abreu et al. (NA50 Collaboration), Phys. Lett. B 450 (1999) 456.  
\bibitem{Abr00} M.C. Abreu et al. (NA50 Collaboration), Phys. Lett. B 477 (2000) 28.   
\bibitem{Abr01} M.C. Abreu et al. (NA50 Collaboration), Phys. Lett. B 521 (2001) 195.  
\bibitem{GRVLO} M. Gl\"uck et al., Z. Phys. C 53 (1992) 127.      
\bibitem{MRS43} A. D. Martin et al., Phys. Rev. D 51 (1995) 4756. 
\bibitem{NA51} A. Baldit et al. (NA51 Collaboration), Phys. Lett. B 332 (1994) 244.     
\bibitem{GRV94} M. Gl\"uck et al., Z. Phys. C 67 (1995) 433.      
\bibitem{Ram03} L. Ramello et al. (NA50 Collaboration), Nucl. Phys. A 715 (2003) 243c.  
\bibitem{Cas03} C. Castanier, Ph.D. Thesis, Universit\'e Blaise Pascal, Aubi\`ere, France, 2003. 
Available at http://cern.ch/NA50/theses.html.
\bibitem{Ale02} B. Alessandro et al., Nucl. Instr. Meth. A 493 (2002) 30.  
\bibitem{Bel97} F. Bellaiche et al., Nucl. Instr. Meth. A 398 (1997) 180.  
\bibitem{Arn98} R. Arnaldi et al., Nucl. Instr. Meth. A 411 (1998) 1.      
\bibitem{Ale04} B. Alessandro et al. (NA50 Collaboration), Eur. Phys. J. C 33 (2004) 31. 
\bibitem{Sha01} R. Shahoyan, Ph.D. Thesis, Instituto Superior T\'ecnico, Lisbon, Portugal, 2001;
Available at http://cern.ch/NA50/theses.html.
\bibitem{Qui02} C. Quintans, Ph.D. Thesis, Instituto Superior T\'ecnico, Lisbon, Portugal, 2002;
Available at http://cern.ch/NA50/theses.html.
\bibitem{San04} H. Santos, Ph.D. Thesis, Instituto Superior T\'ecnico, Lisbon, Portugal, 2004.
Available at http://cern.ch/NA50/theses.html.
\bibitem{DNDETA1} M.C. Abreu et al. (NA50 Collaboration), Phys. Lett. B 530 (2002) 33. 
\bibitem{Kha97} D. Kharzeev et al., Z. Phys. C 74 (1997) 307. 
\bibitem{DNDETA2}  M.C. Abreu et al. (NA50 Collaboration), Phys. Lett. B 530 (2002) 43. 
\bibitem{deJ74} C.W. de Jager et al., Atomic Data and Nuclear Data Tables 14 (1974) 479. 
\bibitem{Trz01} A. Trzcinska et al., Phys. Rev. Lett. 87 (2001) 082501-1. 
\bibitem{NA50-VHI} B. Alessandro et al. (NA50 Collaboration), Charmonium production and nuclear
absorption in p-A interactions at 400 GeV, in preparation. 
\bibitem{Abr03} M.C. Abreu et al. (NA50 Collaboration), Phys. Lett. B 553 (2003) 167. 
\bibitem{Col77} J.C. Collins and D.E. Soper, Phys. Rev. D 16 (1977) 2219. 
\bibitem{Lou95} C. Louren\c co, Ph.D. Thesis, Instituto Superior T\'ecnico, Lisbon, Portugal, 1995;
Available at http://cern.ch/NA38/na38thesis.html.
\bibitem{PYTHIA} T. Sj\"ostrand et al., Computer Phys. Commun. 135 (2001) 238. 
\bibitem{PDFLIB} H. Plothow-Besch, Int. J. Mod. Phys. A 10 (1995) 2901.  
\bibitem{Cap01} L. Capelli, Ph.D. Thesis, Universit\'e Claude Bernard, Lyon, France, 2001;
Available at http://cern.ch/NA50/theses.html.
\bibitem{Sig03} F. Sigaudo, Ph.D. Thesis, Universit\`a di Torino, Turin, Italy, 2003;
Available at http://cern.ch/NA50/theses.html. 
\bibitem{Bor04} G. Borges (for the NA50 Collaboration), J. Phys. G: Nucl. Part. Phys. 30 (2004) S1351. 
\bibitem{NA3} J. Badier et al. (NA3 Collaboration), Z. Phys. C 20 (1983) 101.  
\bibitem{Sch94} G.A. Schuler, Habilitationschrift, Univ. Hamburg, Germany, 1994, 
CERN-TH-7170-94 (hep-ph/9403387). 
\bibitem{GRV98}  M. Gl\"uck et al., Eur. Phys. J. C 5 (1998) 461. 
\bibitem{MRST98} A.D. Martin et al., Eur. Phys. J. C 4 (1998) 463. 
\bibitem{CTEQ5L} H.L. Lai et al. (CTEQ Collaboration), Eur. Phys. J. C 12 (2000) 375, hep-ph/9903282. 

\end{thebibliography}
\end{document}